\pdfoutput=1

\documentclass[a4paper,fleqn]{cas-dc}
\usepackage{bm}

\usepackage[final]{changes}
\usepackage[numbers,sort&compress]{natbib}
\usepackage{color}
\usepackage{amsmath}   
\usepackage{float}
\usepackage{graphicx}  
\usepackage{caption}
\usepackage{subcaption}
\usepackage{booktabs}
\usepackage{array}
\usepackage{tabulary}
\usepackage{listings}
\usepackage{xcolor}
\definecolor{commentgreen}{RGB}{2,112,10}
\definecolor{eminence}{RGB}{108,48,130}
\definecolor{weborange}{RGB}{255,165,0}
\definecolor{frenchplum}{RGB}{129,20,83}
\def\tsc#1{\csdef{#1}{\textsc{\lowercase{#1}}\xspace}}
\tsc{WGM}
\tsc{QE}
\tsc{EP}
\tsc{PMS}
\tsc{BEC}
\tsc{DE}

\begin{document}
	\let\WriteBookmarks\relax
	\def\floatpagepagefraction{1}
	\def\textpagefraction{.001}
	\shortauthors{Hongxuan Zhang et~al.}
	
	\title [mode=title]{Influence of Membrane Characteristics on Efficiency of Vacuum Membrane Distillation: a Lattice Boltzmann Study}

\address[1]{Key Lab of Education Blockchain and Intelligent Technology, Ministry of Education, Guangxi Normal University, Guilin, 541004, China}
\address[2]{Guangxi Key Lab of Multi-Source Information Mining and Security, Guangxi Normal University, Guilin 541004, China}
\address[3]{CAS Key Laboratory of Low-Carbon Conversion Science and Engineering, Shanghai Advanced Research Institute, Chinese Academy of Sciences, Shanghai 201210, China}
\address[4]{PetroChina Shanghai Advanced Materials Research Institute Co.Ltd, Shanghai 201210, China.}

\fntext[equal]{The two authors contribute equally to this work.}

	\author[1,2]{\textcolor{black}{Hongxuan Zhang}\thanks{These authors contributed equally to this work.}}


\author[3,4]{\textcolor{black}{Dian Gong}\thanks{These authors contributed equally to this work.}} 


\author[1,2]{\textcolor{black}{Yiling Zhou}} 


\author[1,2]{\textcolor{black}{Zhangrong Qin}} 
\cormark[1]
\ead{qinzhangrong@gxnu.edu.cn}

\author[1,2]{\textcolor{black}{Binghai Wen}} 
\cormark[1]
\ead{oceanwen@gxnu.edu.cn}

	\begin{abstract}
		With increasing water scarcity, membrane distillation technology has gained widespread attention as an innovative method for seawater desalination. However, existing studies often overlook the influence of membrane characteristics on mass transfer efficiency. This study, based on the lattice Boltzmann method, proposes a model for a novel Poly(tetraethynylpyrene) membrane material to reveal the influence of membrane characteristics on the performance of vacuum membrane distillation. The model considers the factors such as porosity, tortuosity, membrane thickness, pore size, membrane surface wettability and temperature difference on the permeate flux. The results show that the permeate flux increases linearly with the porosity and decreases exponentially with the tortuosity factor. There is an optimal membrane thickness range ($2\,\mu$m) beyond which the permeate flux decreases exponentially. In addition, the permeate flux increases exponentially with increasing temperature difference and pore size. Further analysis of the effect of membrane surface wettability shows that permeate flux increases with increasing hydrophobicity. Finally, the feed temperature and tortuosity factor have the largest effect on permeate flux, followed by membrane thickness and, subsequently, pore size. The model can be further extended to study other configurations of membrane distillation technologies.
	\end{abstract}
		
	\begin{keywords}
Vacuum membrane distillation \sep 
Membrane characteristics \sep 
Mass transfer \sep 
Seawater desalination \sep 
Lattice Boltzmann method
	\end{keywords}

	\maketitle
	
	\section{Introduction}
	\label{sec1}
	Over 70\% of the Earth's surface is covered by water, indicating that there are abundant water resources available for human use. Excluding 97\% of the seawater, of the remaining 3\% of water resources, only a small amount of groundwater and surface water is directly accessible to humans\cite{Chamani2021,suleman2021temperature}. To address the growing problem of water scarcity, desalination and water filtration technologies have been developed focusing on economic and energy efficiency.

In the field of seawater desalination, conventional distillation and membrane desalination are two basic technologies.
Reverse osmosis, an isothermal desalination process, is currently widely used in many applications\cite{qasim2019reverse}. The principle of reverse osmosis is to apply high pressure during operation, causing the solution to pass through the semipermeable membrane, leaving the dissolved salts behind. Although reverse osmosis is considered an effective desalination method, its widespread application is hindered by several limitations, including membrane fouling, degradation and high energy consumption. Membrane distillation (MD), \replaced{as a hybrid process that combines membrane technology with the principles of distillation, offers distinct advantages\cite{el2006framework}.}{on the other hand, offers a promising alternative that has the potential to overcome many of the drawbacks of reverse osmosis.} \replaced{Compared to reverse osmosis, MD operates under non isothermal conditions}{MD is developing rapidly in the field of non-isothermal processes(4). In this process, the feed solution, under the temperature difference}, using a temperature gradient to drive vapor through a semi-permeable membrane, {which then condenses to water on the permeate side\cite{qtaishat2008heat}. \replaced{This mechanism allows MD to operate at atmospheric pressure, avoiding the high pressure conditions required for reverse osmosis, thereby extending membrane life and reducing energy consumption.}{MD has several advantages over reverse osmosis.} \replaced{In addition, MD exhibits an extremely high rejection rate for inorganic compounds, potentially achieving almost complete rejection\cite{khayet2011membranes}}{It has an extremely high rejection rate for inorganic compounds, potentially achieving complete rejection}, \added{and it can use low-grade thermal energy, such as waste heat, geothermal and solar energy, as clean heat sources, further enhancing its sustainability.} MD also offers significant advantages over conventional distillation. Conventional distillation typically requires significant thermal energy for the evaporation and condensation of water, while MD, as a membrane-based separation process, can save approximately 90\% of the energy used in traditional distillation\cite{sholl2016seven}. Moreover, MD operates at lower temperatures, making it suitable for the treatment of heat-sensitive substances. In addition, its system design is more compact, making it ideal for small-scale or distributed applications. In recent years, MD has received increasing attention in various fields, including seawater desalination\cite{zare2018membrane,niknejad2021high}, ethanol recovery\cite{izquierdo2003factors,banat1999modeling}, environmental protection\cite{a2015review}, and radioactive waste treatment \cite{zakrzewska1999concentration,li2023enhanced}. However, the commercialization of MD still faces several challenges, such as a decrease in permeate flux, membrane degradation, and pore wetting\cite{zare2018membrane,niknejad2021mechanically}. However, its unique technical advantages and broad application prospects make it a promising direction for the future advancement of desalination technologies.

The concept of MD was first introduced by Weyl et al.\cite{weyl1967recovery}. Olatunji et al. provided a summary of the heat and mass transfer modeling of the MD configuration\cite{olatunji2018heat}. Shokrollahi et al. indicated that among the factors affecting permeate flux, the order of influence is the flow rate over temperature and the temperature over membrane thickness\cite{shokrollahi2020producing}. Qi et al. found that the effect of vacuum pressure on permeate flux was greater than that of membrane thickness\cite{qi2020numerical}. However, previous studies have not comprehensively considered the effect of different parameters on MD. \added{In a MD system, permeate flux is not only affected by heat and mass transfer processes, but these processes are also controlled by operating conditions and membrane characteristics such as membrane thickness, porosity, tortuosity factor, pore size and pore distribution, which indirectly affect the result. For optimal performance, membranes should be non wetting, have low thermal conductivity, high thermal stability, low \added{resistance to} mass transfer and good chemical resistance. Identifying methods to increase the permeate flux, along with optimization strategies that can guide practical experiments, is a key to improving MD performance. The design of membranes with optimal properties and the research of the ideal combination of parameters is essential\cite{ding2023enhancement,teoh2009membrane}.}

The present work uses a lattice Boltzmann Method (LBM) to model these phenomena, allowing for a more in-depth investigation of the parameters influencing the permeate flux. The advantages of LBM over traditional computational fluid dynamics (CFD) include  better handling of complex flows. \added{MD, as a complex process involving multi-physics coupling, is characterized by the dynamic changes at the liquid-vapor interface and their significant impact on heat and mass transfer processes. Traditional CFD often requires additional interface tracking techniques, such as the volume of fluid (VOF) method, to capture the evolution of phase interfaces when simulating such problems. This not only increases computational complexity but may also lead to reduced accuracy and efficiency. In contrast, LBM based on a mesoscopic kinetic framework, inherently captures phase interfaces and automatically evolves them\cite{shan1993lattice,li2013lattice}, thereby avoiding the technical limitations associated with traditional CFD. Furthermore, LBM demonstrates significant advantages in handling multiphysics coupling problems\cite{kruger2017lattice}, enabling  efficient simultaneous simulation of fluid dynamics, heat transfer, and mass transport processes, while traditional CFD are relatively limited in computational efficiency and coupling capabilities in these areas.
In MD, phase-change phenomena such as liquid evaporation and vapor condensation have a critical impact on system performance. LBM can directly simulate these nucleation phenomena without relying on additional seeding techniques, significantly improving the accuracy and reliability of the simulation\cite{li2015lattice,shan1994simulation}. This seamless handling of physical processes endows LBM with unique advantages in MD simulations, particularly when studying the effects of phase-change phenomena on distillation efficiency and performance. Moreover, MD simulations often require addressing large-scale computational problems, and LBM's parallel computing capabilities enable it to efficiently handle complex simulation tasks at high resolutions, significantly enhancing computational efficiency while maintaining accuracy\cite{qin2024efficient}. Therefore, LBM not only provides an efficient and precise tool for simulating membrane distillation processes but also opens new avenues for research into multi-physics coupling problems. }
\deleted{LBM excels at simulating multiphase flows, providing a more accurate
representation of complicated flow behavior. In addition, in certain scenarios, LBM demonstrates greater numerical stability than computational fluid dynamics, especially when dealing with low Reynolds number flows.}

The MD process has four different configurations: direct contact membrane distillation, sweep gas membrane distillation, air gap membrane distillation, and vacuum membrane distillation (VMD), as shown in Figure 1.

\begin{figure}[htbp]
    \centering
\includegraphics[width=0.45\textwidth]{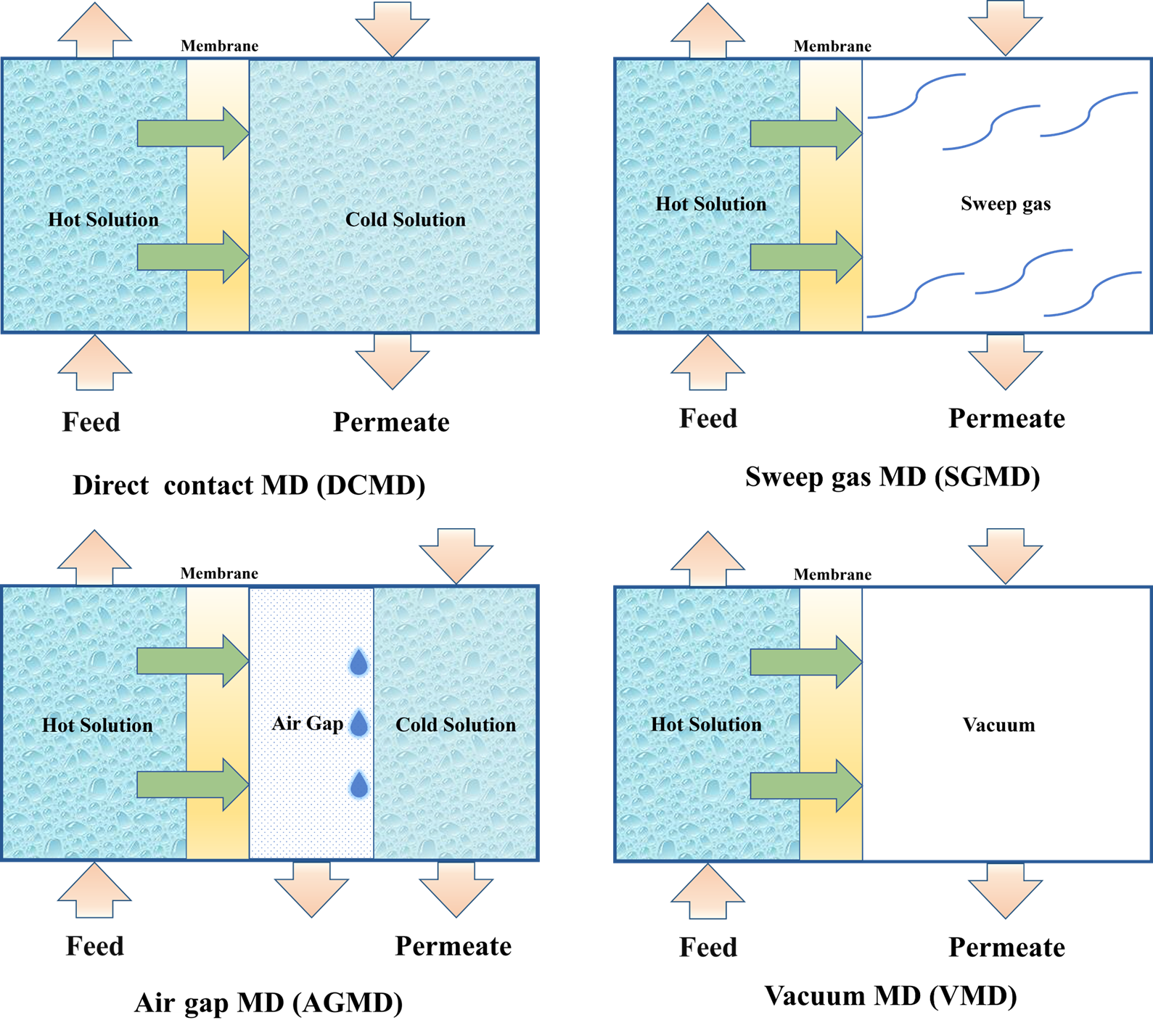}  
\renewcommand{\figurename}{\textnormal{\textbf{Figure}}}
    \caption{\replaced{\rmfamily Basic membrane distillation configurations.}{Basic membrane distillation configurations.}}
    \label{fig:example}  
\end{figure}
A viable approach to improving membrane permeability in MD is to remove air from the membrane pores, either by venting or applying a continuous vacuum on the permeate side, keeping the pressure below the equilibrium vapor pressure using vacuum pumps. This particular configuration, known as VMD, has considerable interest and potential in industrial applications. \added{To elaborate further, firstly, VMD is widely used to remove organic compounds and volatile organic compounds (VOCs)\cite{suk2010development} such as alcohol, toluene and chloroform from aqueous solutions\cite{urtiaga2000kinetic}. Second, VMD plays a key role in concentration processes, including the concentration of sucrose solutions\cite{al2006concentration} and the recovery of aromatic compounds during juice concentration\cite{diban2009vacuum}. Third, it has been shown to be effective in wastewater treatment, especially industrial wastewater containing organic pollutants and dyes\cite{criscuoli2008treatment}. Due to its high efficiency and energy saving characteristics, VMD has become an ideal separation technology in many industrial processes.}

\added{Previous studies have extensively investigated the effects of membrane properties and operating conditions on VMD performance. Although these studies have provided valuable insights, they often rely on simplified models or empirical correlations that may not fully capture the complex interactions of heat and mass transfer in VMD systems. 
In this study, we take a novel approach by using LBM simulations to model heat and mass transfer in the VMD process. In addition, we introduce a new Poly(tetraethynylpyrene) (PTEP) membrane whose performance is predicted by simultaneous modeling and explores the potential of this membrane material to improve the efficiency of VMD. This innovative membrane material, characterized by its unique structural properties, has not been previously investigated in the field of numerical simulation.} \deleted{This study focuses on the use of a novel Poly(tetraethynyl pyrene) (PTEP) membrane for VMD seawater desalination and explores the potential of this membrane material to improve the efficiency of VMD.} The membrane preparation method and the MD test apparatus used in the experiments are described in detail in Section 3. In order to improve the performance of VMD with novel membrane applications and achieve higher permeate flux, a coupled heat and mass transfer model of the VMD system, based on the LBM, was developed and validated with experimental data. \deleted{In order to improve the performance of VMD with novel membrane applications and achieve higher permeate flux, a coupled heat and mass transfer model of the VMD system, based on the LBM, was developed and validated with experimental data.}
	
\section{Theory}
The MD process operates through microporous hydrophobic membranes, where the driving force is maintained by applying a vacuum below the equilibrium vapor pressure on the permeate side and creating a temperature gradient across the membrane. Only water vapor can pass through the hydrophobic pores to the feed side and condense to liquid water on the permeate side.

In VMD, mass transfer is predominantly dominated by the Knudsen diffusion mechanism, which has been confirmed by several research groups\cite{bandini1997vacuum,sarti1993extraction}. The mass flux of the water (\( J \)) is linearly related to its partial pressure difference (\(\Delta P \)) and can be described by the following equation  
\begin{equation}
J = \frac{K_m}{\sqrt{M_w}} \left( P_l - P_v \right),
\end{equation}
where \(M_w\) is the molecular weight of water (kg/mol) in the permeate stream. \(P_l\) and \(P_v\) refer to the downstream pressure and the interfacial partial pressure of water. \(K_m\) is the membrane permeability, described by the following equation
\begin{equation}
K_m=\frac{4 \varepsilon d_p}{3 \replaced{\chi}{\tau} \delta\left(2 \pi R F_T\right)^{1 / 2}},
\end{equation}
which depends on the feed temperature \(F_T\) and membrane characteristics. The properties of the membrane include pore size distribution \(d_p\), membrane thickness \(\delta\), porosity \(\varepsilon\), and tortuosity factor \(\chi\). The universal gas constant is given as \( R \) ($J \cdot mol^{-1} \cdot K^{-1}$).

The tortuosity factor \(\chi\) is a measure of the complexity of the paths through a porous medium, indicating how much longer the actual path is compared to a straight line, which affects the efficiency of mass transport. One of the most successful applied empirical correlations for tortuosity is given by \cite{srisurichan2006mass}
\begin{equation}
\added{\chi}\deleted{\tau}=\frac{2-\varepsilon}{\varepsilon}.
\end{equation}

In the VMD process, heat and mass transfer across the membrane occur simultaneously. In this case, the simple enthalpy balance is
\begin{equation}Q=h_f(T_b-T_I)^I=\sum J\lambda ,\end{equation}
where \(T_b\) is the feed bulk temperature and \(T_l\) is the temperature at the liquid/vapor interface. The latent molar heat of vaporization \(\lambda \) can be reduced using the following formula (valid range: 273 $\sim$ 373 K)
\begin{equation}\lambda(T)=1.7535T+2024.3 ,\end{equation}where \(T\) is expressed in K, and \(\lambda \) in (kJ / kg). \(h_f\) is the heat transfer coefficient in the liquid phase, that can be calculated from Nusselt dimensionless numbers for two hydrodynamic regimes\cite{gryta1997membrane}
\begin{equation}Nu=0.13\operatorname{Re}^{0.64}\operatorname{Pr}^{0.38}(\operatorname{Re}<2100) ,\end{equation}
\begin{equation}Nu=0.023\operatorname{Re}^{0.8}\operatorname{Pr}^{0.33}(\operatorname{Re}>2100),\end{equation}\begin{equation}\mathrm{Re}=\frac{\rho vd_h}{\mu},\end{equation} \begin{equation}\mathrm{Pr}=\frac{C_p\mu}{k_l},\end{equation} where \(v\) is the average velocity of circulation, \(\mu \) is the viscosity,  \(\rho \) is the density and  \(d_h\) is the module's hydraulic diameter, respectively  \(k_l\) is the thermal conductivity of the liquid. The heat transfer coefficients \(h\)  can be estimated by Nusselt number (\(N_u\)) from the following equation
\begin{equation}h=\frac{Nuk_l}{d_h} .\end{equation}

In MD, the membrane surface must be sufficiently hydrophobic to prevent pore wetting. The liquid entry pressure (LEP) refers to the minimum pressure required to force a liquid into the hydrophobic membrane pores and is a key parameter in assessing the membrane's resistance to liquid entry. LEP is influenced by the hydrophobicity of the membrane, the maximum pore size, and the presence of organic solutes. For example, LEP decreases linearly with increasing ethanol content in the solution\cite{gostoli1989separation}. Based on Franken et al.'s work\cite{franken1987wetting}, LEP can be estimated using the following equation
\begin{equation}LEP=\Delta P=P_f-P_p=\frac{-2B\gamma_l\cos\theta}{r_{\max}} ,\end{equation}where \(P_f\) and \(P_p\) are the hydraulic pressure on the feed and permeate side, \(B\) is the geometric pore coefficient (equal to 1 for cylindrical pores), \(\gamma_{l}\) is liquid surface tension, \(\theta \)  is contact angle and \(r_{\max}\) is the maximum pore size.
	
\section{Preparation of PTEP membranes}
The porous structure within the PTEP membrane is fabricated by a rapid solid-state Glaser coupling reaction of the powdery monomer tetraethynylpyrene (TEP) under heated conditions. During the coupling reaction, TEP undergoes rapid volumetric expansion, resulting in a foam-like porous structure. Figure 2 shows the scanning electron microscope (SEM) images of the porous structure of the fabricated PTEP membrane.
\begin{figure}[htbp]
    \centering
\includegraphics[width=0.45\textwidth]{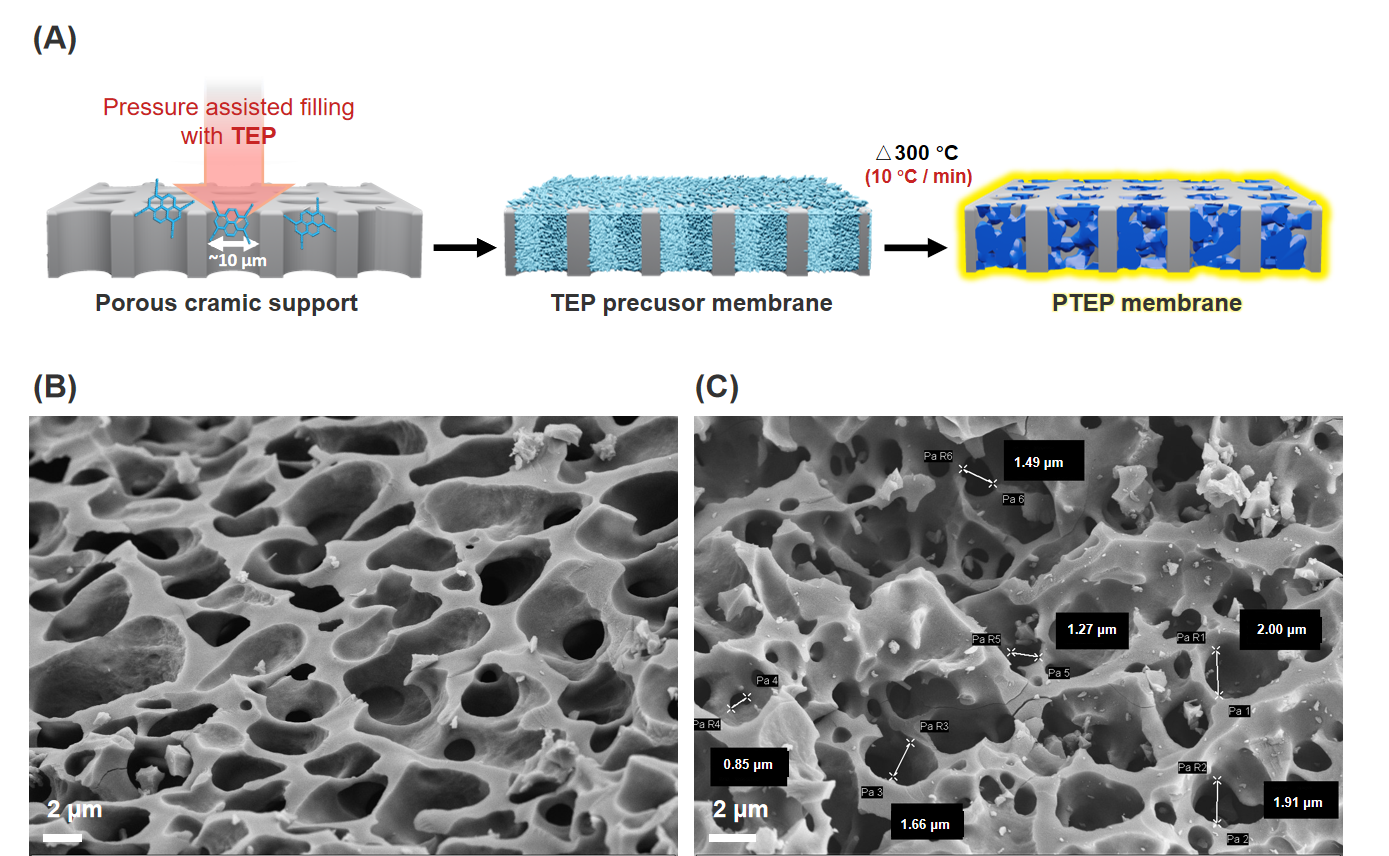}  
    \renewcommand{\figurename}{\textnormal{\textbf{Figure}}}
    \caption{\rmfamily(A) Schematic diagram of the PTEP membrane preparation process; (B) SEM image of the porous structure on the PTEP membrane surface; (C) Pore size distribution of the PTEP membrane surface ($\sim$1-$2\,\mu$m)}  
    \label{fig:example}  
\end{figure}

 The membrane is prepared through the following steps: 1 mg (0.0034 mmol) TEP was dispersed in 300 mL methanol and sonicated for 30 minutes. The TEP suspension was vacuum-filtered onto a porous ceramic support to form a precursor TEP membrane, and was dried in air at 40 °C. The TEP membrane was then heated in a tubular furnace at 10 °C/min to 300 °C, triggering the in-situ Glaser coupling reaction. The reaction completed in 5 minutes, resulting in a PTEP membrane with a distinctive interconnected microporous structure. The prepared PTEP membrane was allowed to dry at room temperature for 5 hours before undergoing performance testing for MD. The schematic of the MD test equipment used in the experiment is presented in Figure 3. \added{The monomer 1,3,6,8-tetraethynylpyrene (TEP,98.0\%) was purchased from Shanghai Tengqian. Methanol (>98.0\%) used to disperse TEP were purchased from Adamas. The sodium chloride (NaCl, 99.9\%) used to configure simulated seawater was purchased from Greagent. The multi-temperature tubular furnace was purchased from Shanghai Bona Electric Furnace Co., Ltd.}

\begin{figure}[htbp]
    \centering
\includegraphics[width=0.45\textwidth]{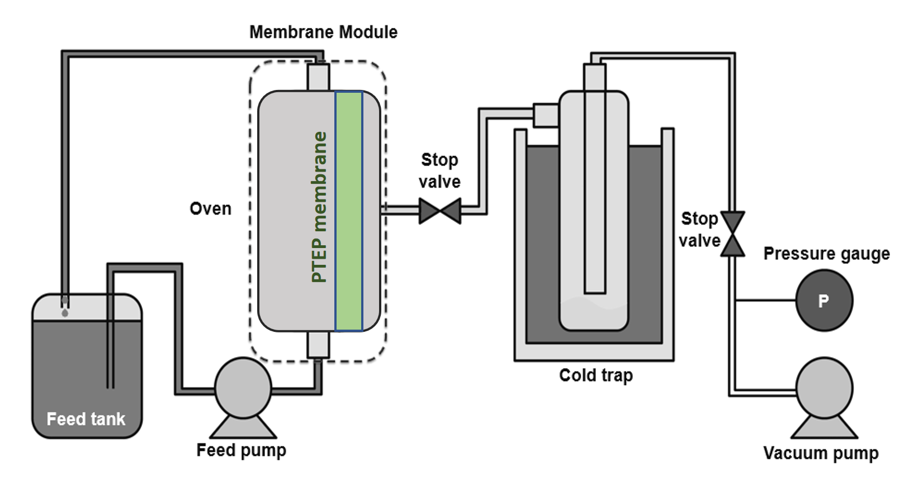}  
\renewcommand{\figurename}{\textnormal{\textbf{Figure}}}
    \caption{\replaced{\rmfamily Schematic of vacuum membrane distillation.}{VMD process configurations.}}  
    \label{fig:example}  
\end{figure}

\added{
The PTEP membrane introduced in this paper, as a completely new membrane material, has not yet entered into real-world applications, but it has unique properties in theory. First, as an alkyne-based conjugated framework material, PTEP has an atomically thick two-dimensional topology, a periodic and uniform distribution of in-plane pores, and graphene-like surface properties, which make it ideal for efficient vapor transport and separation in VMD\cite{qiu2019graphynes}. Its crystalline structure and reliable chemical/mechanical strength indicate great potential for efficient desalination, and its uniform pore distribution and hydrophobicity effectively prevent liquid penetration while allowing vapor permeation, which is critical for VMD performance\cite{yin2022ultrafast}. In addition, the PTEP membranes were constructed under mild conditions via the Glaser-Hay cross-coupling reaction, which further supports the feasibility of their large-scale production in practical applications. This method ensures the robustness and homogeneity of the membrane structure, which is critical for stable performance in VMD systems\cite{gong2024alkadiyne}.
}

\section{Numerical Model}
Understanding the potential heat and mass transfer mechanisms within the VMD system is crucial to developing an accurate model. The VMD system consists of three parts: the feed side, the membrane side, and the permeate side. All three parts involve heat and mass transfer phenomena. Due to the complexity of these physical systems, significant computational resources are required to simulate the various parameters involved. In addition, the governing equations should be coupled, requiring the simultaneous solution of each dependent variable at a given point to accurately explain the phenomena. In this context, the LBM is particularly suited to meet these requirements.

\subsection{Lattice Boltzmann Method}
The lattice Boltzmann Method originated from the concept of cellular automaton and kinetic theory\cite{aidun2010lattice,chen1998lattice}, the intrinsic mesoscopic properties make LBM outstanding in modeling complex fluid systems involving interfacial dynamics and phase transitions. The numerical simulation of multiphase flows is one of the most successful applications in the LBM field\cite{li2016lattice,chen2014critical,huang2015multiphase}. In this study, we selected the multiple relaxation time LBM to obtain better numerical stability and computational accuracy, which can be concisely expressed as\cite{lallemand2000theory}
\begin{equation}f_{i}\left(\boldsymbol{x}+\boldsymbol{e}_{i} \delta_{t}, t+\delta_{t}\right)-f_{i}(\boldsymbol{x}, t)=-\mathbf{M}^{-1}\left[\mathbf{S}(\mathbf{m}-\mathbf{m}^{\mathrm{eq}})+\delta_{t} \mathbf{F}_{m}\right],\end{equation}
where \(f_i(\boldsymbol{x}, t)\) is the particle distribution function at time \(t\) and lattice site \(\boldsymbol{x}\), moving along the direction defined by the discrete velocity vector \(\boldsymbol{e}_{i}\) with \(i=0, ..., N\), \(\boldsymbol{\mathbf{m}}\) and \(\mathbf{m}^{\mathrm{eq}}\) represent the velocity moments of the distribution functions and their equilibria, respectively, \(\boldsymbol{\mathbf{M}}\) is a transformation matrix that linearly transforms the distribution functions to the velocity moments, \(\mathbf{m=Mf}\); and \(\mathbf{f=M}^{-1}\mathbf{m} ,\) where \(\mathbf{f}=(f_0,f_1,...,f_N)^{\mathrm{T}}\). For the two-dimensional nine-velocity model(D2Q9) model on a square lattice, \(N\) is equal to 8, the velocity moments are \(\mathbf{m}=(\rho,e,\varepsilon,j_x,q_x,j_y,q_y,p_{xx},p_{xy})^\mathrm{T}\). The conserved moments are the density \(\rho=\sum_{i}f_{i}\) and the flow momentum \(\boldsymbol{J}=(j_x,j_y)=\rho\boldsymbol{u}=\sum_i{\boldsymbol{e}_i}f_{i}\). The equilibria of non-conserved moments depend only on the conserved moments: \(e^{\mathrm{(eq)}}=-2 \rho+\frac{3}{\rho}(j_x^2+j_y^2)\), \(\varepsilon^{\mathrm{(eq)}}=\rho-\frac{3}{\rho}(j_x^2+j_y^2)\), \(q_x^{\mathrm{(eq)}}=-j_x\), \(q_y^{\mathrm{(eq)}}=j_y\), \(p_{xx}^{\mathrm{(eq)}}=\frac{1}{\rho}(j_x^2-j_y^2)\), \(p_{xy}^{\mathrm{(eq)}}=\frac{1}{\rho}(j_xj_y)\). The diagonal relaxation matrix has the non-negative relaxation rates, \(\mathbf{S}=\mathrm{diag}(1,s_{_e},s_{_\varepsilon},1,s_{_q},1,s_{_q},s_{_\nu},s_{_\nu})\), in which \(s_e=1.64\), \(s_\varepsilon=1.54\), \(s_q=1.7\)\cite{mccracken2005multiple}. The shear viscosity is\(\quad v=\frac13(\frac1{s_{\nu}}-\frac12)\); in the present study, \(s_{\nu}=\frac1\tau \), and the relaxation time is \(\tau=0.7\). The evolution of LBM includes two essential steps, namely collision and streaming. The force term \(\mathbf{F}_{m}\) is incorporated the external force into the lattice Boltzmann equation, and this study uses the exact difference method proposed by Kupershtokh et al.\cite{kupershtokh2009equations}, which is simply equal to the difference of the equilibrium moments before and after the force acting on the fluid during a time step 
\begin{equation}
\mathbf{F}_{m} = \boldsymbol{\mathrm{m}}^{\mathrm{(eq)}}(\rho, \boldsymbol{u} + \Delta \boldsymbol{u}) - \boldsymbol{\mathrm{m}}^{\mathrm{(eq)}}(\rho, \boldsymbol{u}),\end{equation}
where \(\Delta \boldsymbol{u}=\delta_t \boldsymbol{F} / \rho\).
Consequently, the macroscopic velocity is defined as the average momentum before and after the collision: \(\boldsymbol{v}=\boldsymbol{u}+\delta_t\boldsymbol{F}/(2\rho)\)\cite{shan1995multicomponent}.

\subsection{Chemical-potential multiphase model}
The chemical potential is the partial molar Gibbs free energy at constant pressure. Both the Onsager and Stefan–Maxwell formulations of irreversible thermodynamics recognize that the chemical potential gradient is the driving force for isothermal mass transport. The movement of molecules from higher to lower chemical potential is accompanied by a release of free energy, and the chemical or phase equilibrium is achieved at the minimum free energy. For a nonideal fluid system, following the classical capillarity theory of van der Waals, the free energy functional within a gradient-squared approximation is written as\cite{lallemand2000theory,mccracken2005multiple,wen2017chemical,wen2020chemical,d1992generalized} \begin{equation}\Psi=\int\left[\psi(\rho)+\frac{\kappa}{2}\left|\nabla\rho\right|^2\right]d\boldsymbol{x}\end{equation} where \(\psi\) is the bulk free energy density at a given temperature, \(\kappa\) is the surface tension coefficient, and the square of the gradient term gives the free energy contribution from density gradients in an inhomogeneous system. The chemical potential can be derived from the free energy density functional \cite{jamet2002theory} 
 \begin{equation}\mu=\psi^{\prime}(\rho)-\kappa\nabla^2\rho.\end{equation}
  The nonlocal pressure is related to free energy by
  \begin{equation}p=p_0-\kappa\rho\nabla^2\rho^{}-\frac{\kappa}{2}{\left|\nabla\rho\right|}^2\end{equation}
with the general EOS defined by the free energy density
\begin{equation}p_0=\rho\psi^{\prime}(\rho)-\psi(\rho).\end{equation}
With respect to the ideal gas pressure, the nonideal force can be evaluated by a chemical potential
 \begin{equation}\boldsymbol{F}=-\rho\nabla\mu+c_s^2\nabla\rho .\end{equation}
 Because the derivation is within thermodynamics, it is expected that the phase transition induced by the nonideal force theoretically satisfies the thermodynamic consistency and Galilean invariance, which have been confirmed by numerical simulations of static and dynamic fluids. 
 
The Peng–Robinson (PR) EOS is superior in predicting liquid densities:
\begin{equation}p_0=\frac{\rho RT}{1-b\rho}-\frac{a\alpha(T)\rho^2}{1+2b\rho-b^2\rho^2},\end{equation} where \(a\) is the attraction parameter, and \(b\) is the volume correction. In this simulations, the parameters are given by \(R\)=1,  \(a\)=2/49, \(b\)=2/21. \(\alpha(T)=[1+(0.37464+1.54226\omega-0.26992\omega^2)\times(1-\sqrt{T/T_c})]^2\) is the temperature function and the acentric factor \(\omega\) is assigned to 0.344 for simulating the water/vapor system. Its chemical potential is
\begin{align}
\mu &= R T \ln\frac{\rho}{1\!-\!b\rho}
-\frac{a\alpha(T)}{2\sqrt{2}b}
\ln\frac{\sqrt{2}-1\!+\!b\rho}{\sqrt{2}+1\!-\!b\rho} \notag \\
&\quad + \frac{R T}{1\!-\!b\rho}
-\frac{a\alpha(T)\rho}{1\!+\!2b\rho\!-\!b^{2}\rho^{2}}
-\kappa\nabla^{2}\rho\;.
\end{align}
To make the numerical results closer to the actual physical properties, we define the reduced variables \(T_r=T/T_c\) and \(\rho_r=\rho/\rho_c\), in which \(T_c\) is the critical temperature and \(\rho_{c}\) is the critical density.

A proportional coefficient \(k\) is introduced to decouple the computational mesh from the moment space\cite{wen2020chemical}, and relate the length units of the two spaces, \(\hat{\delta x}=k\delta x .\) Here, the quantities in the mesh space are marked with a superscript. According to dimensional analysis, the chemical potential in the mesh space can be calculated by the following method\cite{wen2017chemical,wen2020chemical} \begin{equation}\hat{\mu}=k^2\psi'(\rho)-\hat{\kappa}\hat{\nabla}^2\rho .\end{equation} This method greatly improves the stability of the chemical-potential multiphase model, and the transformation maintains mathematical equivalence without loss of accuracy.

The chemical potential can effectively describe the wettability of solid interfaces\cite{swift1995lattice}. For solid boundaries of curved surfaces, the density of the solid node layers should be assigned based on adjacent nodes\cite{liu2023contact}\begin{equation}\rho(\boldsymbol{x}_s)=\frac{\sum_i\omega_i\rho(\boldsymbol{x}_s+\boldsymbol{e}_i\delta_t)s_\omega}{\sum_i\omega_is_\omega},\end{equation}where \(\boldsymbol{x}_s+\boldsymbol{e}_i\delta_t\) represents adjacent nodes, and \(s_\omega\) is a switching function. For the first layer of solid nodes, \(s_\omega\)=1 when \(\boldsymbol{x}_s+\boldsymbol{e}_i\delta_t\) is a fluid node; for the second or third layer of nodes which depend on whether the fourth-order accuracy or the sixth-order accuracy central difference method is used, \(s_\omega\)=1 when \(\boldsymbol{x}_s+\boldsymbol{e}_i\delta_t\) is in the first or second layer, respectively; otherwise, \(s_\omega\)=0. When dealing with a horizontal solid boundary, the above equation can be simplified to a weighted average algorithm based on the densities of adjacent node layers.

\subsection{Heat transfer}
In this model, the temperature evolves according to the standard diffusion-advection equation\cite{lallemand2003theory}\begin{equation}\partial_tT+\boldsymbol{u}\cdot\nabla T=\zeta\Delta T+(\gamma-1)c_{s 0}^2\nabla\cdot\boldsymbol{u} ,\end{equation}where \(\zeta\) is the thermal diffusivity, \(\gamma\) is the ratio of specific heats and \(c_{s 0}\) is the isothermal sound velocity, respectively. The linear stability analysis shows that so long as \(\gamma\) is not too far away from 1, the model is stable. The advection-convection equation is solved by the following finite-difference equation \begin{equation}T(\boldsymbol{x},t+\delta_t)-T(\boldsymbol{x},t)=-\boldsymbol{u}\cdot\nabla T+\zeta\Delta T+(\gamma-1)c_{s0}^2\nabla\cdot\boldsymbol{u} .\end{equation}

Using the membrane evaporation model based on LBM, we analyze the impact of each factor on the permeate flux and rank them according to their significance. Specifically, we assess how variations in porosity \(\varepsilon,\) tortuosity factor \(\chi\), membrane thickness \(\delta,\) pore size \(d_{p},\) temperature difference and contact angle influence the permeate flux, determining the extent of their effects and establishing a hierarchy of their contributions to the overall performance of the MD process.

\subsection{Simulation Details}
In this work, we applied the chemical-potential multiphase lattice Boltzmann model, which is thermodynamically consistent and can simulate large density ratios. Together with the chemical-potential boundary condition, it is easy to model the surface wettability with the linearly adjustable contact angles. The present simulations are performed in the two-dimensional plane with fluids confined within the computational domain \(0\leq x\leq L_{x}\) and \(0\leq x\leq L_{y}\). Here, \(L_{y}\) is a fixed value of 400, which is converted to the real length of 40 \(\mu m\), and \(L_{x}\) is adjusted according to different porosities and pore widths. Under normal circumstances, the value of \(L_{x}\) is 200 or 240. The periodic boundary condition is applied in the x direction to close the system. Solid nodes representing the membrane are also present in the central part of the flow field. The boundaries of the membrane pores are represented by two parallel sine functions, and the physical shape of the pores is adjusted by changing the amplitude and period of the sine functions. The liquid on the lower surface of the membrane represents the feed side, while the flow field above the membrane represents the permeate side, where the evaporated water vapor will condense to the top of the field.  

The membrane surface adopts a hydrophobic contact angle 150° to prevent the pore wetting, which reduces the permeate flux because the solution below the membrane infiltrates the pores. For the flow field below the membrane \((y\leq175),\) our simulation starts from the equilibrium state of two-phase coexistence at the temperature \(T_{s1}=0.610 T_{c}\) corresponding to the coexistence densities \(\rho_V\approx0.012\) and \(\rho_L\approx8.668\). For the flow field above the membrane \((y>175),\) the simulation also starts from the equilibrium state of gas-liquid coexistence at the temperature \(T_{s2}=0.665 T_{c}\) corresponding to the coexistence densities \(\rho_V\approx0.033\) and \(\rho_L\approx8.552\). The critical temperature and densities are solved by the equation of Maxwell equal-area construction. In the computational domain, the parts of \(75\leq y<125\) and \(370\leq y<400\) are initially set as liquid, and the rest is set as vapor. The relaxation time \(\tau\) is set to 0.7. 

According to the novel PTEP membrane mentioned in Section 3, the parameters of this model are set as follows: \(\varepsilon=0.25 , \replaced{\chi}{\tau}=1.4 , \delta=5 \mu m ,\quad d_p=2 \mu m\) and the contact angle of the membrane surface is \(150^{\circ}\). In the following simulations, the parameters remain unchanged unless otherwise specified. \added{And in this study, all experiments are conducted under conditions close to a complete vacuum (with vacuum pressure approaching -100.1 kPa).}

\section{Results and discussion}
	The VMD simulation model developed in this study was used to evaluate the performance of the VMD process. Specifically, this model investigated how critical membrane properties, including porosity, pore size, wettability, membrane thickness and tortuosity factor, as well as operating conditions such as temperature difference, influence membrane transport behavior. By investigating the role of these factors, the model provides insight into optimizing the efficiency of the VMD process and identifies potential ways to improve permeate flux and separation performance for practical applications.
\subsection{Model Validation}
To validate the accuracy of the model, the simulation results were compared with published experimental data. We selected experimental data of the novel PTEP membrane with different feed temperatures as comparison objects to ensure the applicability of the model under different conditions. As shown in Figure 4, the simulation results are in close agreement with the experimental data, with a maximum discrepancy of 6.37 \(\mathrm{L\cdot m^{-2}\cdot h^{-1}}\)
(LMH) observed at a feed temperature of 40°C, while the differences remain below 2 LMH in other cases. Moreover, the total relative error between the simulated and experimental data is \(1.18\%\), indicating that the developed model effectively captures the heat and mass transfer phenomena in the MD process with high reliability and precision. Therefore, this model can be used for further parameter studies and optimization designs, providing theoretical support for the practical application of VMD technology.
\begin{figure}[htbp]
    \centering
\includegraphics[width=0.45\textwidth]{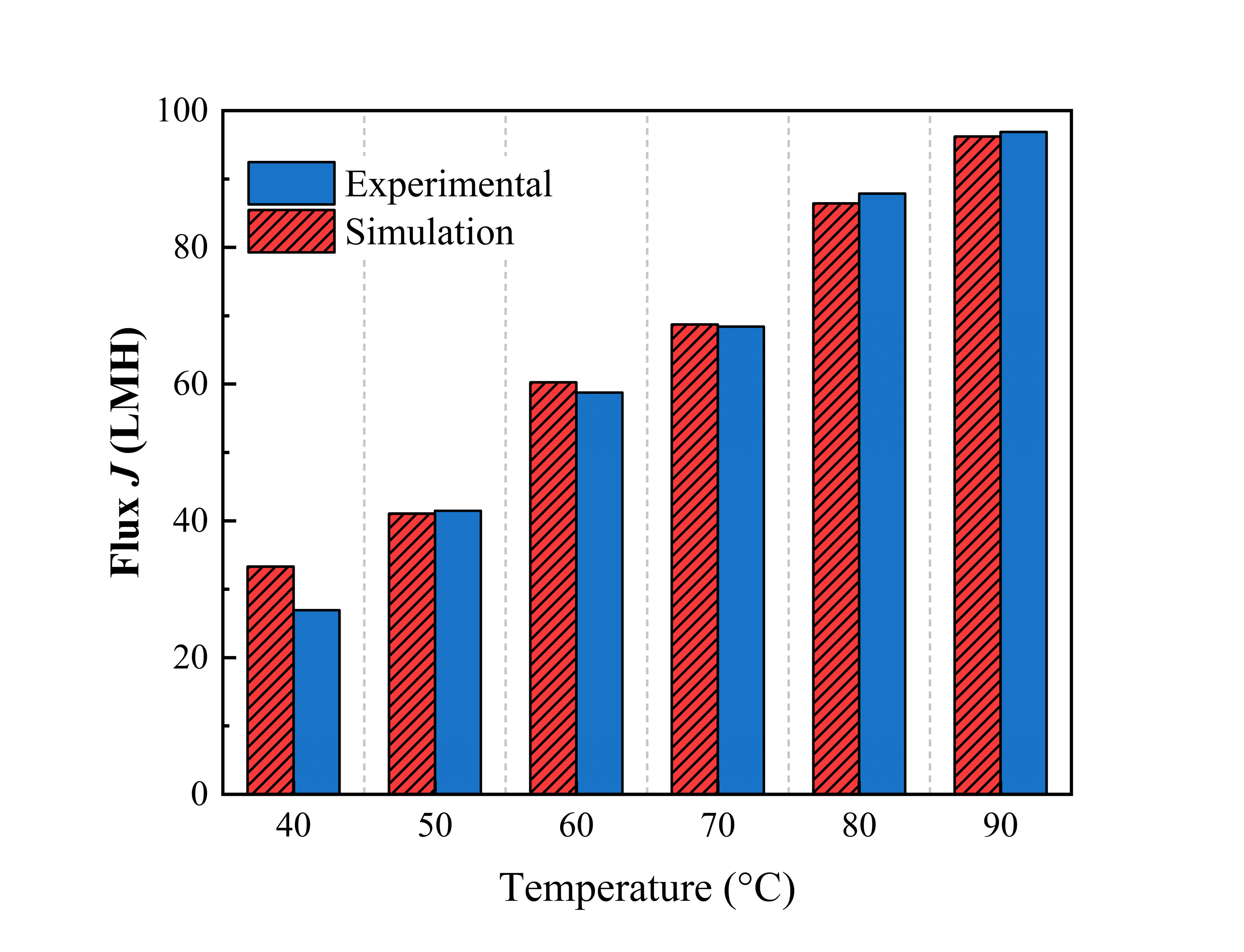}  
\renewcommand{\figurename}{\textnormal{\textbf{Figure}}}
    \caption{\replaced{\rmfamily The results of the present model are highly consistent with the experimental data.}{The results of the present model are highly consistent with the experimental data.}}  
    \label{fig:example}  
\end{figure}
\subsection{Effect of membrane porosity}
Porosity  $\varepsilon$ is a critical parameter that significantly affects permeate flux in the VMD process. Defined as the ratio of the pore volume to the total membrane volume, porosity reflects the overall pore structure within the membrane and directly affects its transport efficiency. Figures 5 and 6 illustrate the relationship between membrane porosity and VMD performance. Figure 5 illustrates the relationship between porosity  $\varepsilon$ and permeate flux $J.$ As porosity increases, permeate flux also increases, suggesting a direct relationship. In Figure 6, the linear fit (shown as a straight line) emphasizes this positive correlation, with higher porosity corresponding to higher permeate flux. This trend occurs because membranes with greater porosity provide a larger effective surface area for evaporation, thereby increasing the transport of water vapor across the membrane and increasing the overall permeate flux. \added{Besides, according to Darcy's Law, in porous media, the permeate flux $J$ of a fluid is directly proportional to the porosity $\varepsilon$, \( J \propto \varepsilon \). Therefore, an increase in porosity $\varepsilon$ directly leads directly to a linear growth in permeate flux $J$. This positive correlation is not only supported by experimental data but also aligns with mass transfer theory, indicating that enhancing membrane porosity is an effective strategy for optimizing permeate flux in the VMD process.}
\begin{figure}[htbp]
    \centering
\includegraphics[width=0.45\textwidth]{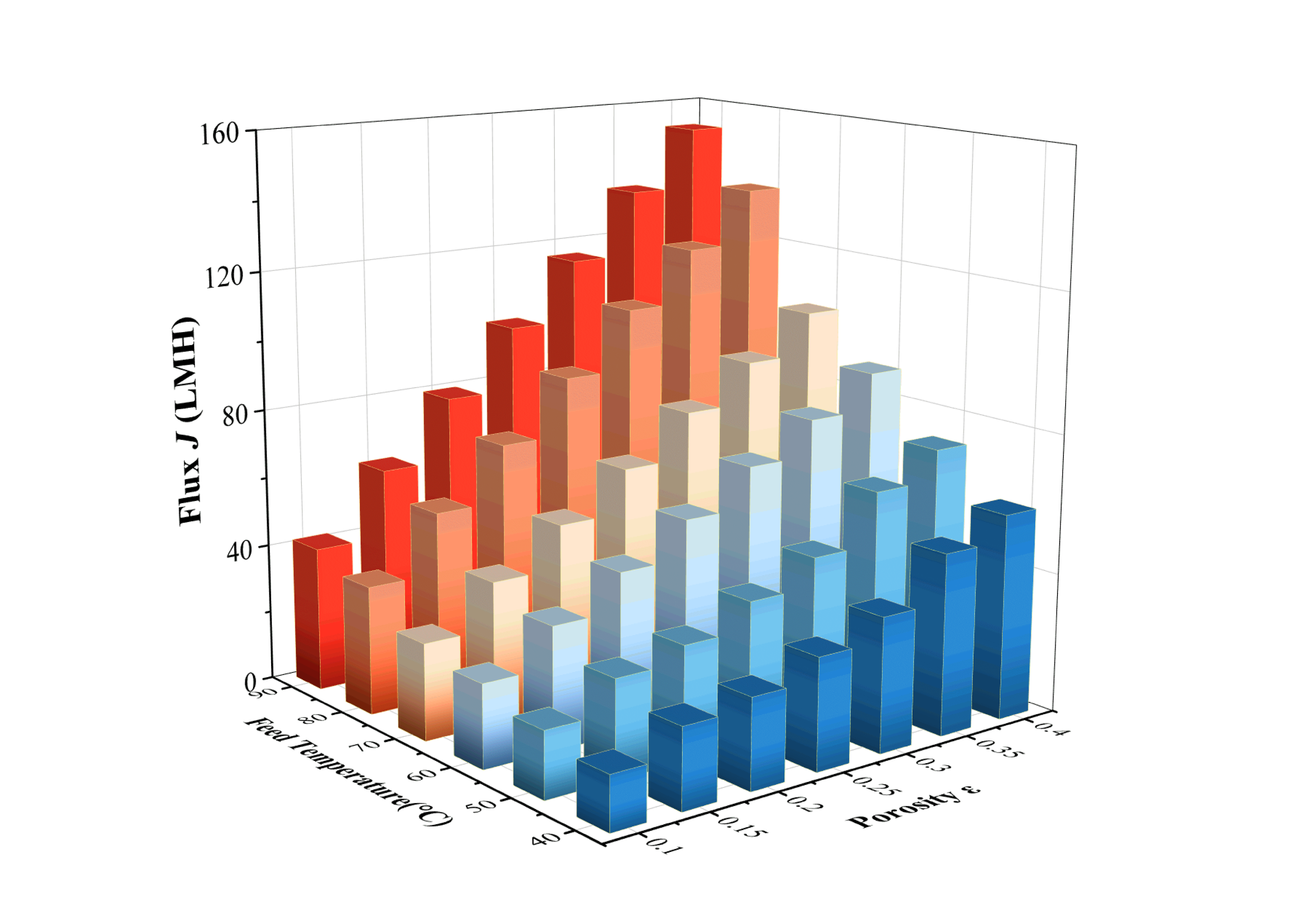}
\renewcommand{\figurename}{\textnormal{\textbf{Figure}}}
    \caption{\replaced{\rmfamily Effect of the membrane porosity $\varepsilon$ on permeate flux $J$ in the VMD process.}{ Effect of the membrane porosity $\varepsilon$ on permeate flux $J$ in the VMD process.} }  
    \label{fig:example}  
\end{figure}
\begin{figure}[htbp]
    \centering
\includegraphics[width=0.45\textwidth]{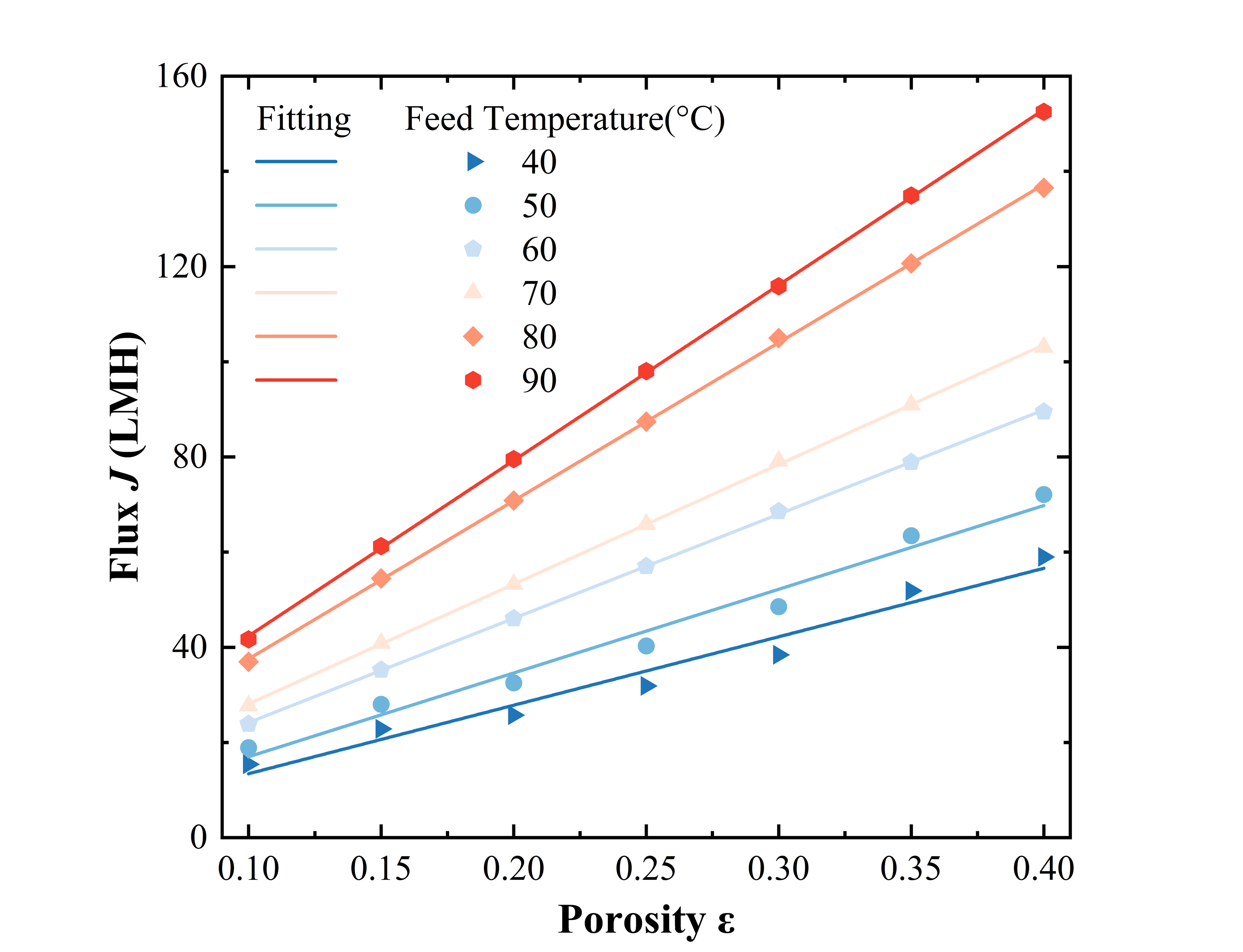}  
\renewcommand{\figurename}{\textnormal{\textbf{Figure}}}
    \caption{\replaced{\rmfamily Linear relationship between membrane porosity $\varepsilon$ and permeate flux $J$ (fitted results).}{Linear relationship between membrane porosity $\varepsilon$ and permeate flux $J$ (fitted results).}}  
    \label{fig:example}  
\end{figure}
\subsection{Effect of membrane tortuosity factor}
Understanding transport phenomena within porous membranes relies heavily on the parameter of tortuosity \replaced{$\chi$}{$\tau$}. It quantifies the complexity of pore paths within the membrane by representing the ratio of the actual length of a pore to its apparent (vertical) length. In essence, the tortuosity factor provides insight into the ease with which fluids can navigate the structure of a membrane. In simple terms, a pore that runs directly through the membrane has a tortuosity factor of 1.0. A more complex path results in a factor greater than 1.0. To predict transmembrane flux, tortuosity \replaced{$\chi$}{$\tau$} is typically assumed to be 2\cite{khayet2004characterization}. 

In Figure 7, panels (a), (b) and (c) illustrate the effect of increasing membrane
tortuosity on permeate flux at feed temperatures of 50°C, 70°C, and 90°C, respectively.
\begin{figure}[htbp]
    \centering
\includegraphics[width=0.45\textwidth]{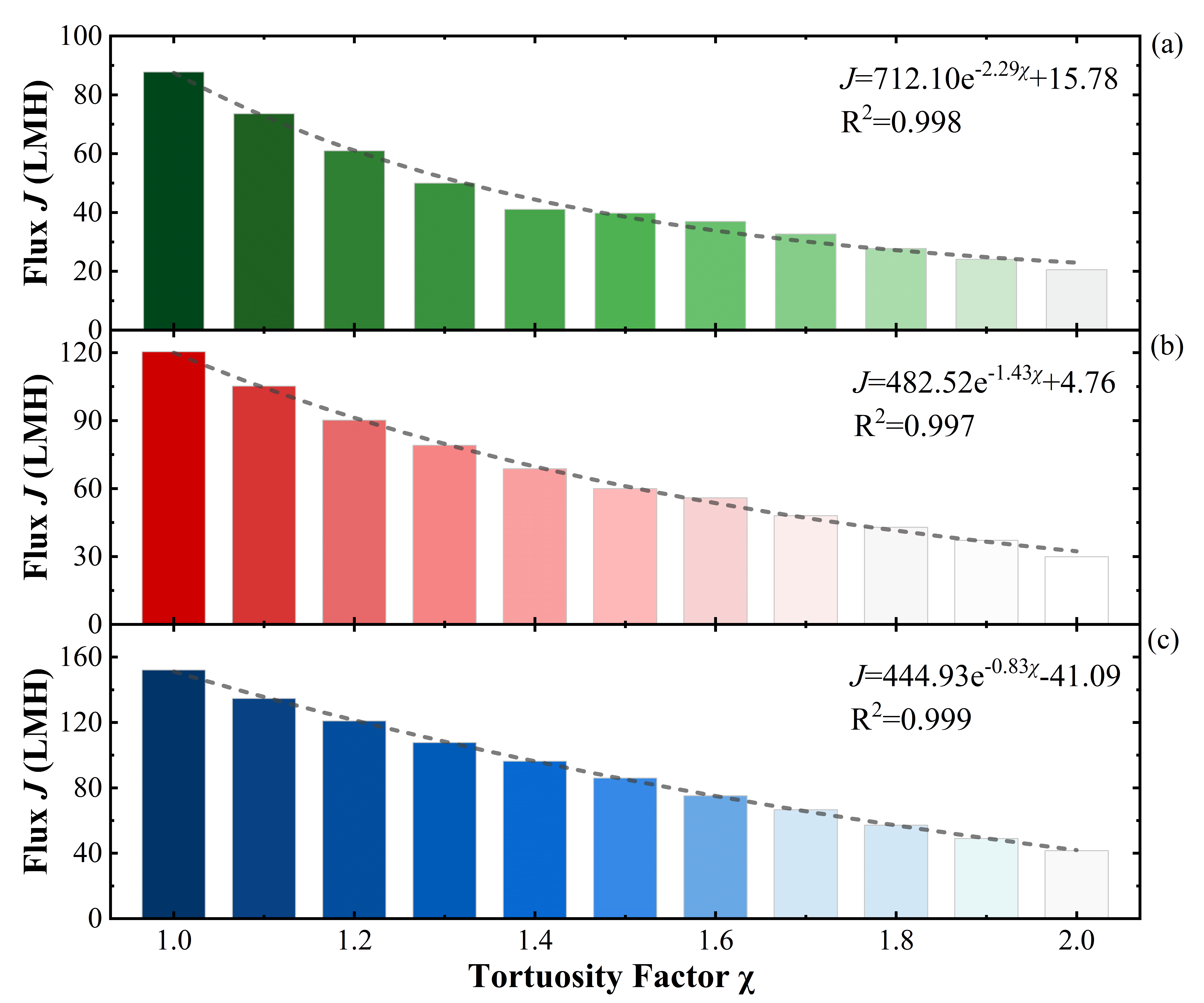}  
\renewcommand{\figurename}{\textnormal{\textbf{Figure}}}
 \caption{\rmfamily Tortuosity \(\chi\) and permeate flux \(J\) under varying feed temperatures. }  
    \label{fig:example}  
\end{figure}

\added{In this section, we validated the selection of the tortuosity factor $\chi$ within the range of 1.0 - 2.0 through a sensitivity analysis. The results indicate that this range adequately reflects the physical characteristics of the system and has a minimal impact on the qualitative trends of the model results, thus supporting our parameter choice.}
As can be seen, as the tortuosity $\chi$ increases, the permeate flux decreases in an exponential decay pattern, indicating that higher tortuosity values correspond to reduced permeate flux. \replaced{This phenomenon can be explained by Fick's law, which states that the flux of a substance is proportional to the concentration gradient. As the transport distance increases, the concentration gradient decreases, leading to a reduction in flux. The specific form of this reduction depends on the physical context and the model setup. In this model, the reduction in flux follows an exponential relationship, as the diffusion resistance gradually increases with path length, causing the concentration gradient to decrease at an accelerating rate, which in turn leads to an exponential decay of the flux. The curvature of the membrane pores increases the flow resistance, requiring molecules to overcome more friction and viscous resistance, further exacerbating the flux reduction. In addition, the curved pores also affect the efficiency of molecular diffusion by changing the direction of fluid flow, making the diffusion process less efficient and further reducing flux.}{This phenomenon occurs because membrane pores are rarely straight, and water vapor molecules must travel along these tortuous paths. As the tortuosity increases, the distance that molecules must travel through the membrane increases, resulting in greater resistance and consequently lower permeate flux.}
\subsection{Effect of membrane thickness}
The performance of separation processes is significantly affected by the thickness of the membrane. Membranes with greater thickness tend to have improved structural integrity and mechanical strength, which can be beneficial in high pressure applications. However, increased thickness can also result in increased mass transfer resistance, which in turn results in decreased permeate flux. As a result, there is frequently a trade-off between the membrane's mechanical properties and its permeability. \added{The effective thickness of the PTEP film is determined by the penetration depth of the reactive monomer TEP into the support during preparation ($5\,\mu$m). This thickness is challenging to increase in practice due to the requirements of the solid-phase Glaser coupling reaction, which necessitates instantaneous heating. Excessive film thickness leads to uneven heat conduction, and finally only TEP near the surface can react to form a film, resulting in an invariant effective film thickness of ($5\,\mu$m). The mechanical strength of the PTEP membrane is sufficient for VMD applications. In our current experiments, the membrane demonstrated stable operation for over 80 hours under a pressure of approximately -100 kPa without significant loss in separation performance. This operational duration is expected to improve with further advancements in our ongoing research. }

In many applications, achieving an optimal balance between durability and efficiency requires the optimization of membrane thickness. A thinner membrane may allow for higher permeate flux rates, but it is also more susceptible to fouling and physical damage. By carefully designing and selecting the optimal membrane thickness, researchers and engineers can enhance the overall performance of membrane systems and ensure that they meet the specific requirements of applications such as water treatment, desalination, and industrial separations.
 \begin{figure}[htbp]
    \centering
\includegraphics[width=0.45\textwidth]{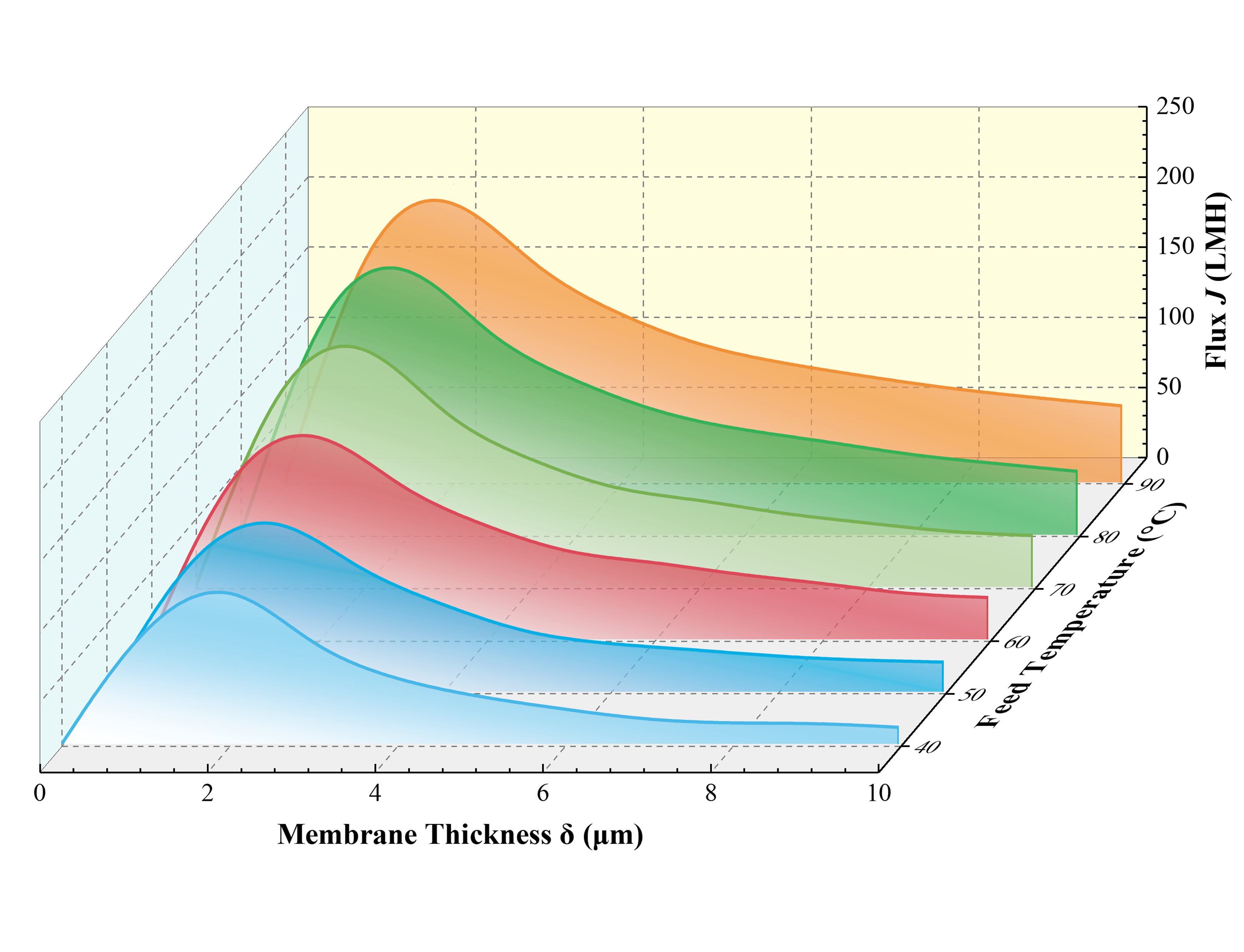}  
\renewcommand{\figurename}{\textnormal{\textbf{Figure}}}
    \caption{\rmfamily Membrane thickness \(\delta\) and permeate flux \(J\)  under varying feed temperatures.}  
    \label{fig:example}  
\end{figure}

As shown in Figure 8, for the specified set of operating and design conditions, the model predicts the membrane thickness that maximizes permeate flux. This phenomenon may be the result of competition between the water vapor diffusion path across the membrane and the driving force. As the membrane thickness decreases, the diffusion path for water vapor is reduced, promoting an increase in permeate flux. However, the driving force also decreases (the temperature difference across the membrane decreases), which suppresses the permeate flux. After the maximum permeate flux is reached, the permeate flux decreases with increasing membrane thickness according to a power law relationship.

This power-law relationship indicates that the influence of the membrane microstructure and physical properties on permeate flux is complex. Once the membrane thickness reaches a certain value, the increase in diffusion distance significantly slows down the transport of water vapor, while excessive membrane thickness increases the resistance to vapor flow, leading to a significant decrease in permeate flux. Therefore, it is critical to determine the optimum membrane thickness when designing and optimizing membranes to achieve efficient water vapor transport and overall system performance.

\subsection{Effect of pore size}
The pore size of a membrane directly affects the selectivity and permeate flux of the membrane. A larger pore size allows a greater amount of vapor to pass through, thereby increasing the permeate flux. Using pores that are too large can result in a decrease in the selectivity of the membrane for solutes, which can ultimately lead to a decrease in product quality. It is therefore imperative that the pore size is optimized in a manner that is appropriate to the context of the membrane design.

\begin{figure}[htbp]
    \centering
\includegraphics[width=0.45\textwidth]{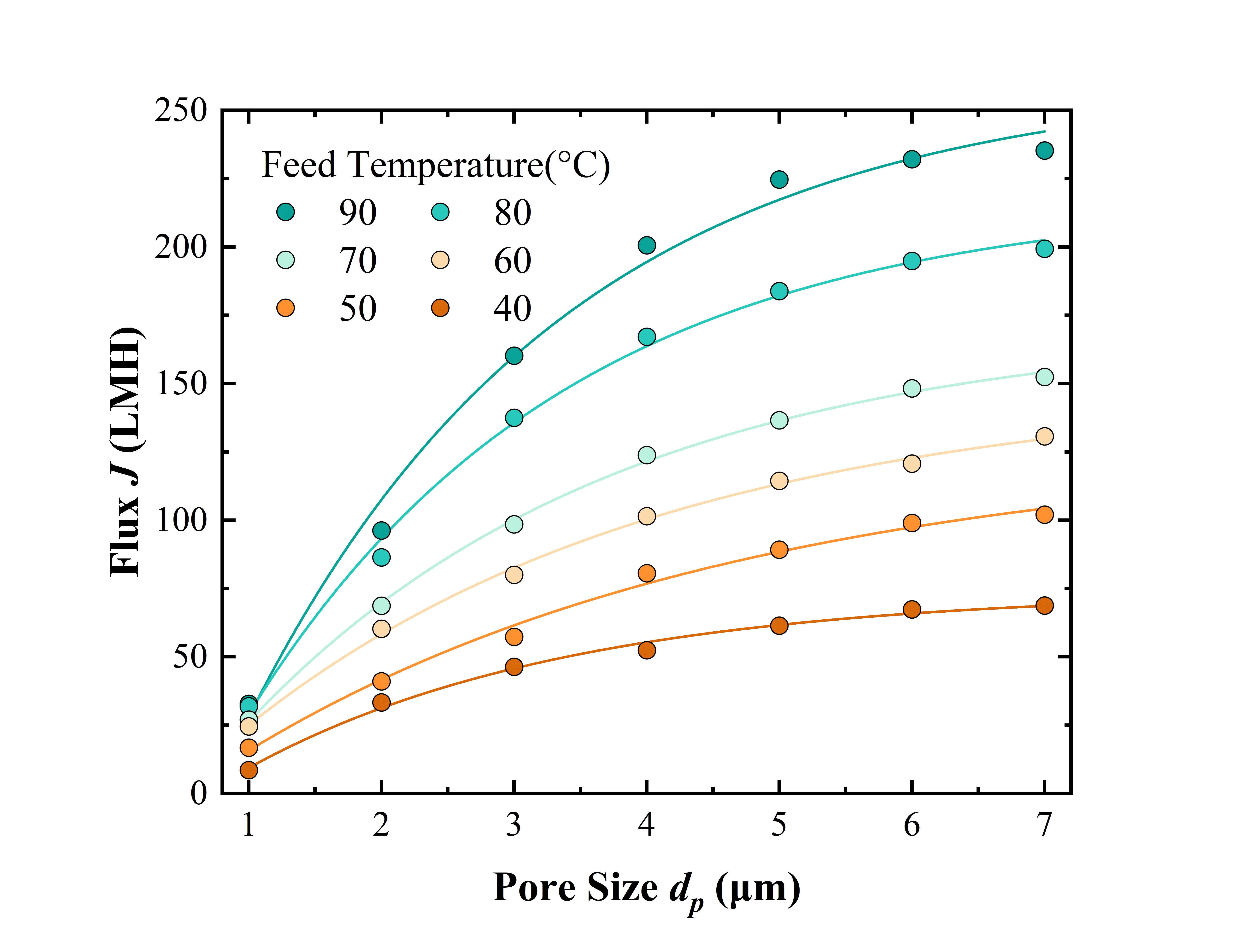}  
\renewcommand{\figurename}{\textnormal{\textbf{Figure}}}
    \caption{\rmfamily Effect of pore size \(d_p\) on permeate flux \(J\) at different feed temperature.}  
    \label{fig:example}  
\end{figure}
The effect of pore size $d_p$ on permeate flux is shown in Figure 9. It is evident that as the pore size increases from 1 $\mu m$ to 7 $\mu m$, the permeate flux increases dramatically. \deleted{This trend can be attributed to the reduced mass transfer resistance resulting from the increased pore size, which allows the vapor to pass through the membrane more easily. It is noteworthy that the slope of the curve gradually decreases, indicating that the rate of increase in permeate flux decreases as the pore size continues to increase.}
\added{This phenomenon is primarily due to the significant reduction in mass transfer resistance with increasing pore size. A larger pore size provides a wider passageway for vapor molecules, reducing friction and viscous resistance within the membrane, making it easier for molecules to pass through the membrane and significantly increasing the permeate flux. However, as the pore size continues to increase, the rate of permeate flux increase gradually slows down, which is reflected in the gradual decrease in the slope of the curve in Figure 9. This is because when the pore size becomes sufficiently large, permeate flow resistance is no longer the primary limiting factor and diffusion effects gradually become more significant. Although increasing the pore size still reduces the mass transfer resistance, the effect on permeate flux improvement diminishes so that the permeate flux increase no longer follows a linear relationship, but instead exhibits exponential growth and eventually levels off. This change reflects the diminishing returns of pore size on permeate flux, where the permeate flux increase becomes smaller as the pore size increases.} The line in Figure 9 represents the fitted curve showing an exponential growth relationship between pore size $d_p$ and permeate flux $J$,with the specific function prototype shown as \begin{equation}J=A_1\exp(-d / t_1)+y_0\end{equation} and the parameters listed in Table 1. 

\begin{table}[width=1\linewidth,cols=5,pos=h]
\caption{\replaced{\rmfamily Fitting table: pore size and permeate flux.}{Fitting table: pore size and permeate flux.}}\label{tbl1}
\begin{tabular*}{\tblwidth}{@{} CCCCC@{} }
\toprule
\rmfamily Feed Temperature & \(A_1\)& \(t_1\)& \(y_0\)& \(R^2\)\\
\midrule
\rmfamily40 & \rmfamily-97.741 & \rmfamily24.227 & \rmfamily74.102 &\rmfamily0.991 \\
\rmfamily50 & \rmfamily-144.942 & \rmfamily37.645 & \rmfamily126.882 &\rmfamily0.989 \\
\rmfamily60 & \rmfamily-168.932 & \rmfamily31.897 & \rmfamily148.541 &\rmfamily0.996\\
\rmfamily70 & \rmfamily-207.099 & \rmfamily28.051 & \rmfamily171.279 &\rmfamily0.998\\
\rmfamily80 & \rmfamily-286.032 & \rmfamily24.128 & \rmfamily218.201 &\rmfamily0.994\\
\rmfamily90 & \rmfamily-353.952 & \rmfamily24.019 & \rmfamily261.455 &\rmfamily0.988\\
\bottomrule
\end{tabular*}
\end{table}
Based on the simulation conclusions, wider and straighter pore channels can theoretically enhance the efficiency of MD. However, this serves merely as a theoretical guideline. In practical membrane fabrication, it is often challenging to simultaneously control multiple physical parameters. For example, increasing the width of pores can lead to greater interconnectivity between pores, thereby reducing the tortuosity. In contrast, too short pore channels may not prevent wetting of the feed solution. Therefore, determining the optimal pore size should not only consider the impacts of the physical parameters discussed in this study but also account for the chemical stability (susceptibility to hydrolysis or degradation), and mechanical strength (e.g., porous structures with excessively large pores may be prone to collapse). Particularly when employing novel membrane materials and structures that lack a substantial experimental basis, the influence of pore size parameters should be regarded as an important theoretical reference rather than an absolute directive. Based on Figure 10, we predict that $5\,\mu$m may be the best choice if the change in permeate flux is not significant.
\subsection{Effect of temperature difference}
In VMD, the feed temperature has a significant effect on the permeate flux. In this section, the permeate side temperature is fixed at 25°C, and as the feed temperature increases, the temperature gradient across the membrane widens, increasing the vapor pressure difference, the kinetic energy of the water molecules, and decreasing the mass transfer resistance, which allows more water vapor to pass through the membrane, resulting in an increase in permeate flux. Experimental result shows a positive correlation between feed temperature and permeate flux as the temperature difference increases from 13°C to 61°C. Figure 10 illustrates the trend of permeate flux with feed temperature, showing a significant increase in permeate flux with increasing temperature, and the curve fits well with the Srichards function as\begin{equation}y=a*(1+\left(d-1)*\exp(-k*\left(x-xc)\right))\right.^{\frac1{1-d}},\end{equation} indicating a predictable relationship. The parameters of the Srichards function are shown in Table 2.

\begin{figure}[htbp]
    \centering
\includegraphics[width=0.45\textwidth]{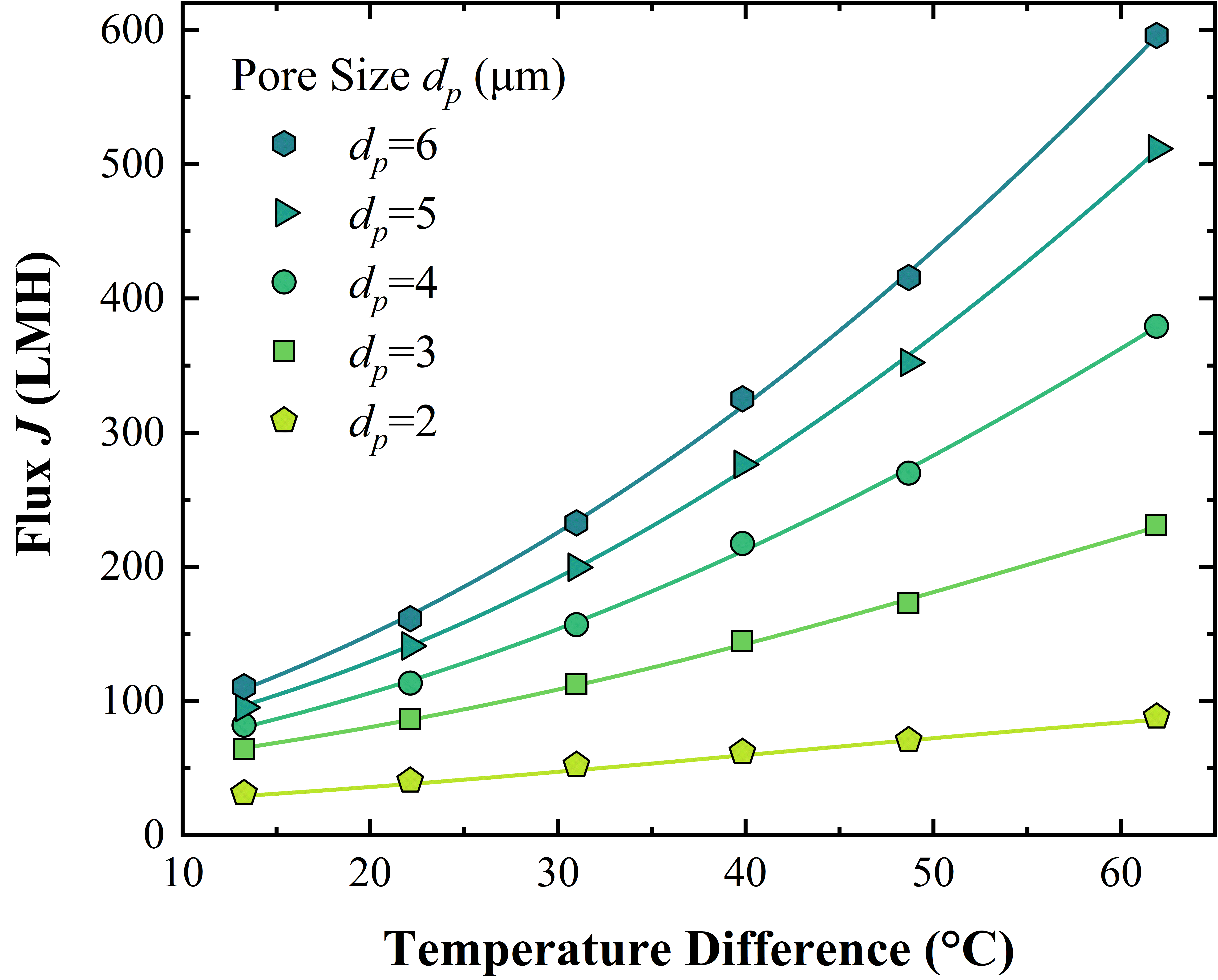}  
\renewcommand{\figurename}{\textnormal{\textbf{Figure}}}
    \caption{\rmfamily The correlation between temperature difference and permeate flux \(J\) at varying pore sizes. }  
    \label{fig:example}  
\end{figure}

\begin{table}[width=1\linewidth,cols=6,pos=h]
\caption{\replaced{\rmfamily Fitting table: temperature difference and permeate flux.}{Fitting table: temperature difference and permeate flux.}}\label{tbl2}
\begin{tabular*}{\tblwidth}{@{} CCCCCC@{} }
\toprule
\rmfamily Pore Size & \(a\)& \(x_c\)& \(d\)&\(k\)& \(R^2\)\\
\midrule
\rmfamily2 & \rmfamily121.289 & \rmfamily41.226 & \rmfamily2.064 & \rmfamily0.042 &\rmfamily0.957 \\
\rmfamily3 & \rmfamily408.200 & \rmfamily56.961 & \rmfamily126.882& \rmfamily0.041 &0.997 \\
\rmfamily4 & \rmfamily1614.989 & \rmfamily88.974 & \rmfamily148.541& \rmfamily0.016 &0.997\\
\rmfamily5 & \rmfamily2032.261 & \rmfamily85.697 & \rmfamily171.279& \rmfamily0.020 &0.998\\
\rmfamily6 & \rmfamily5612.823 & \rmfamily132.051 & \rmfamily218.201& \rmfamily0.008 &0.999\\
\bottomrule
\end{tabular*}
\end{table}

\subsection{Effect of contact angle}
The wettability of the membrane surface is an important property that affects membrane performance because it determines the behavior of liquids on the membrane surface. Good wettability helps improve the contact and penetration of water molecules, thereby enhancing permeate flux. However, if the membrane surface is overly hydrophilic, it can increase the adhesion of contaminants, resulting in a decrease in membrane lifetime and efficiency.

The effect of contact angle on permeate flux is shown in Figure 11. 
\replaced{When the contact angle exceeds 90°, the membrane surface becomes more hydrophobic, meaning that the interaction between water molecules and the membrane surface is weaker. A higher contact angle helps prevent pore wetting because water molecules are more likely to roll off the membrane surface rather than stick to it, keeping the pores dry and preventing liquid water from blocking vapor flow. This ensures that vapor can easily pass through the membrane, maintaining high permeation efficiency. At the same time, greater hydrophobicity is associated with a stronger driving force for water vapor movement, allowing vapor to diffuse more effectively through the membrane surface. Therefore, membranes with a higher contact angle exhibit better vapor permeation efficiency during the diffusion process.}
{When the contact angle exceeds 90°, a higher contact angle indicates greater surface hydrophobicity, which increases the membrane's ability to repel liquid water and prevent pore wetting, thereby maintaining efficient vapor transport. In addition, a larger contact angle correlates with a stronger driving force for water vapor movement.} As shown in the latter part of Figure 11, when the membrane material is predominantly hydrophobic, an increased contact angle results in a higher permeate flux. However, when the contact angle falls below 90°, the permeate flux is significantly higher compared to hydrophobic membranes. This does not indicate that hydrophilic membranes perform better under VMD conditions. The observed phenomenon is mainly due to pore wetting, where some liquid water enters the membrane pores, leading to unexpectedly high final results. This phenomenon is discussed further in Figure 12.
\begin{figure}[htbp]
    \centering
\includegraphics[width=0.45\textwidth]{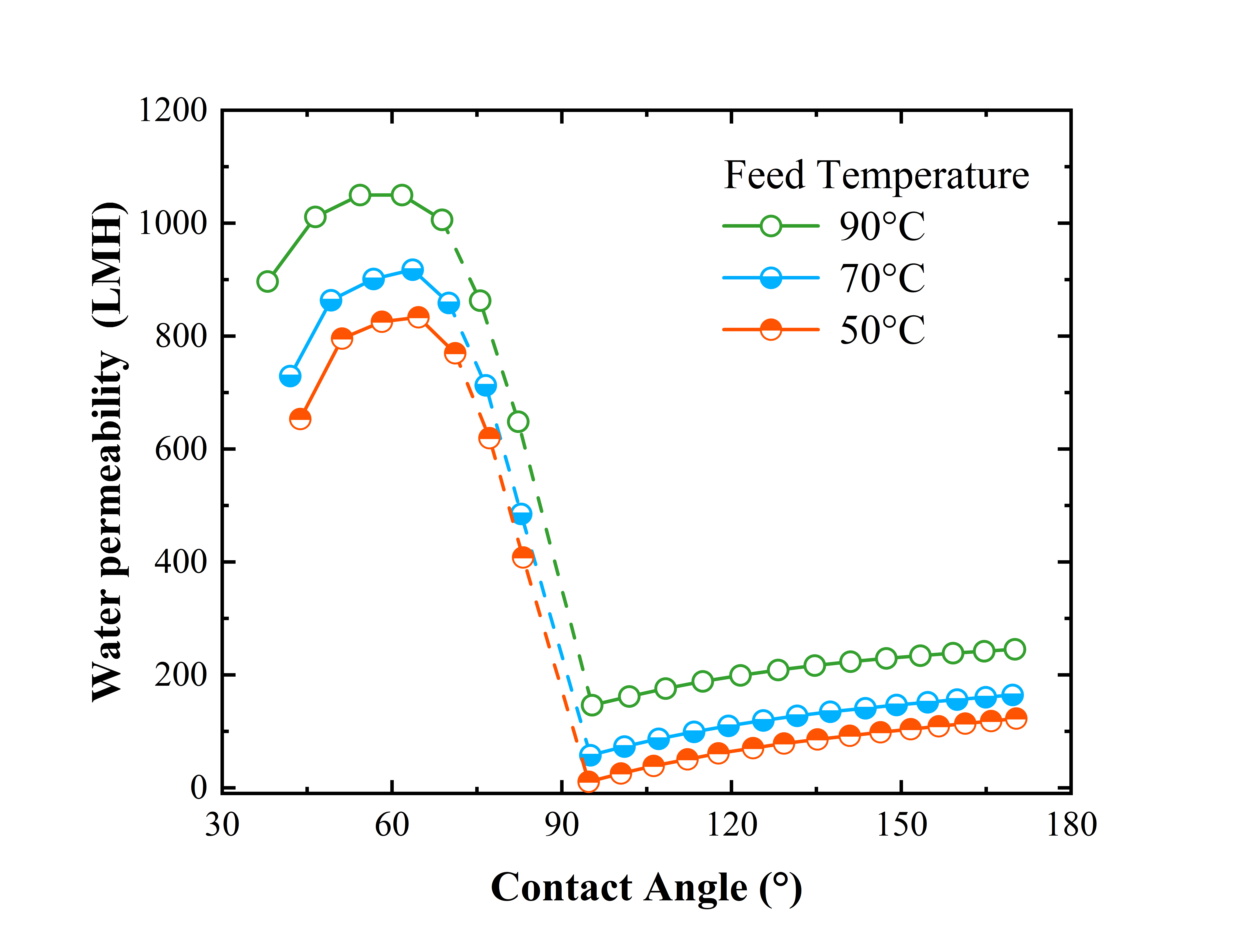}  
\renewcommand{\figurename}{\textnormal{\textbf{Figure}}}
    \caption{\replaced{\rmfamily Effect of membrane hydrophilicity and hydrophobicity on water flux at different feed temperatures.}
    {Effect of membrane hydrophilicity and hydrophobicity on water flux at different feed temperatures.}}  
    \label{fig:example}  
\end{figure}

Figure 12 shows the evolution of four different contact angles (50°, 70°, 110°, 130°). At contact angles of 110° and 130°, the liquid water under the membrane, influenced by the contact angle, does not enter the membrane pores but remains in the area below the pores. Conversely, at contact angles of 50° and 70°, the wetting pressure exceeds the liquid entry pressure, resulting in pore wetting that is significantly faster at a contact angle of 50°.

\begin{figure}[htbp]
    \centering
\includegraphics[width=0.45\textwidth]{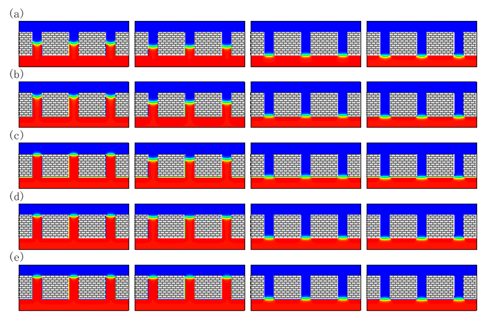}  
\renewcommand{\figurename}{\textnormal{\textbf{Figure}}}
    \caption{\replaced{\rmfamily The evolution of the solution under the membrane at different wettability conditions. From left to right, the contact angles are 50°, 70°, 110°, and 130°, while (a) to (e) represent different evolution times at $1.34 \times 10^{-5}$ s, $2.68 \times 10^{-5}$ s, $4.02 \times 10^{-5}$, $5.36 \times 10^{-5}$ s, and $6.67 \times 10^{-5}$ s, respectively.}{The evolution of the solution under the membrane at different wettability conditions. From left to right, the contact angles are 50°, 70°, 110°, and 130°, while (a) to (e) represent different evolution times at $1.34 \times 10^{-5}$ s, $2.68 \times 10^{-5}$ s, $4.02 \times 10^{-5}$, $5.36 \times 10^{-5}$ s, and $6.67 \times 10^{-5}$ s, respectively.}}  
    \label{fig:example}  
\end{figure}

\subsection{The influence of various factors on permeate flux}
Certain factors can affect permeate flux in opposite directions, such as membrane thickness and pore size, which can have competing effects on the permeation rate. Numerous parameters affect permeate flux in the VMD process, making it challenging to isolate and adjust just one in real-world applications. Therefore, exploring how these parameters interact and identifying dominant relationships is critical for optimizing performance. To this end, we present 3D response surface plots that illustrate the simultaneous effects of selected operating parameters on permeate flux, as shown in Figure 13 to Figure 15.

Figure 13 shows the response surface of permeate flux as a function of tortuosity and feed temperature, while holding other parameters constant. The surface plot clearly illustrates the combined effect of tortuosity and feed temperature on the permeation process, which ranges from a minimum of 16.79 to a maximum of 152.09 LMH. The shape of the surface indicates that the optimum permeate mass is obtained when the tortuosity factor is low and the feed temperature is high. From this figure, it is clear that the influence of tortuosity factor and feed temperature on permeate flux is comparable, with each parameter having a significant effect on permeation efficiency.
\begin{figure}[htbp]
    \centering
\includegraphics[width=0.5\textwidth]{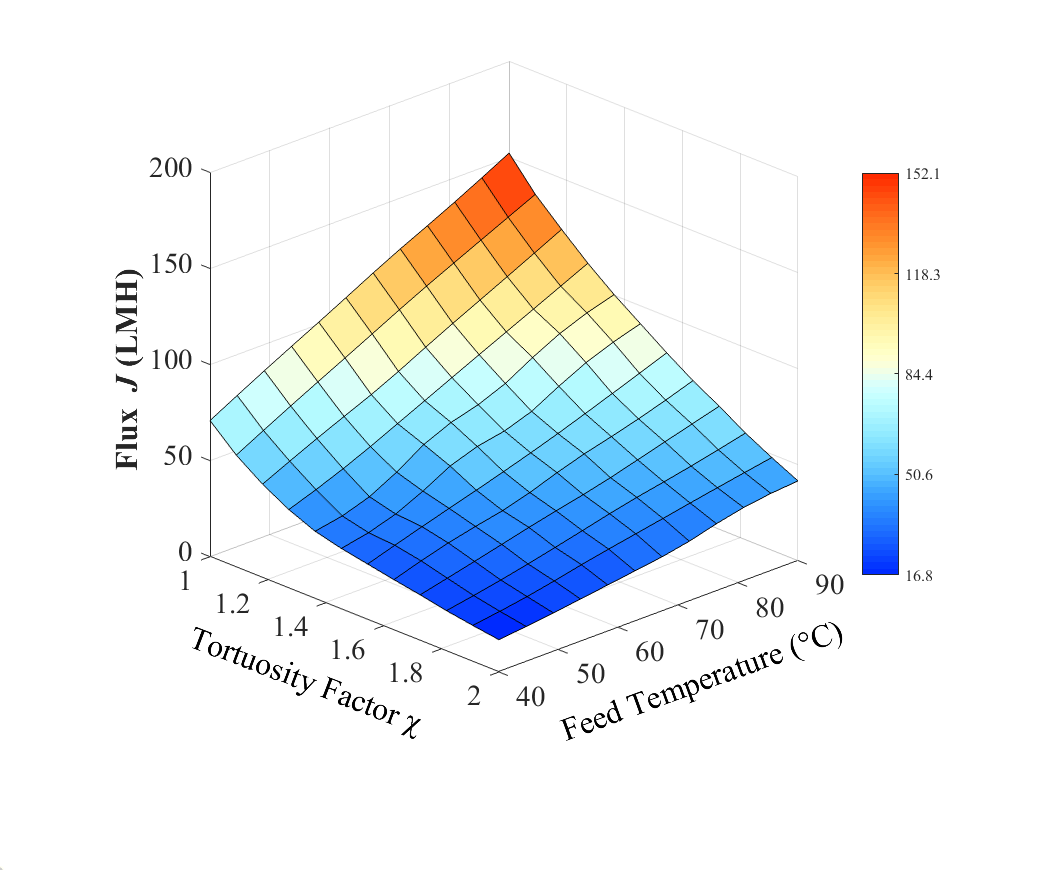}  
\renewcommand{\figurename}{\textnormal{\textbf{Figure}}}
    \caption{\replaced{\rmfamily Response surface plots of permeate flux as a function of feed temperature and tortuosity factor.}{Response surface plots of temperature polarization as a function of feed temperature and tortuosity factor.}}  
    \label{fig:example}  
\end{figure}

Figure 14 aims to investigate the relationship between permeation quality, membrane thickness, and tortuosity, and to visualize how these two factors influence permeation efficiency. In this scenario, the membrane thickness is set between 3 and 9 \(\mu m\), while the tortuosity ranges from 1 to 2, with all other parameters held constant. The surface plot shows that tortuosity has a much stronger effect on permeation quality compared to membrane thickness, indicating that designing membranes with minimized tortuosity may be more advantageous for improving permeation quality in practical applications.
\begin{figure}[htbp]
    \centering
\includegraphics[width=0.45\textwidth]{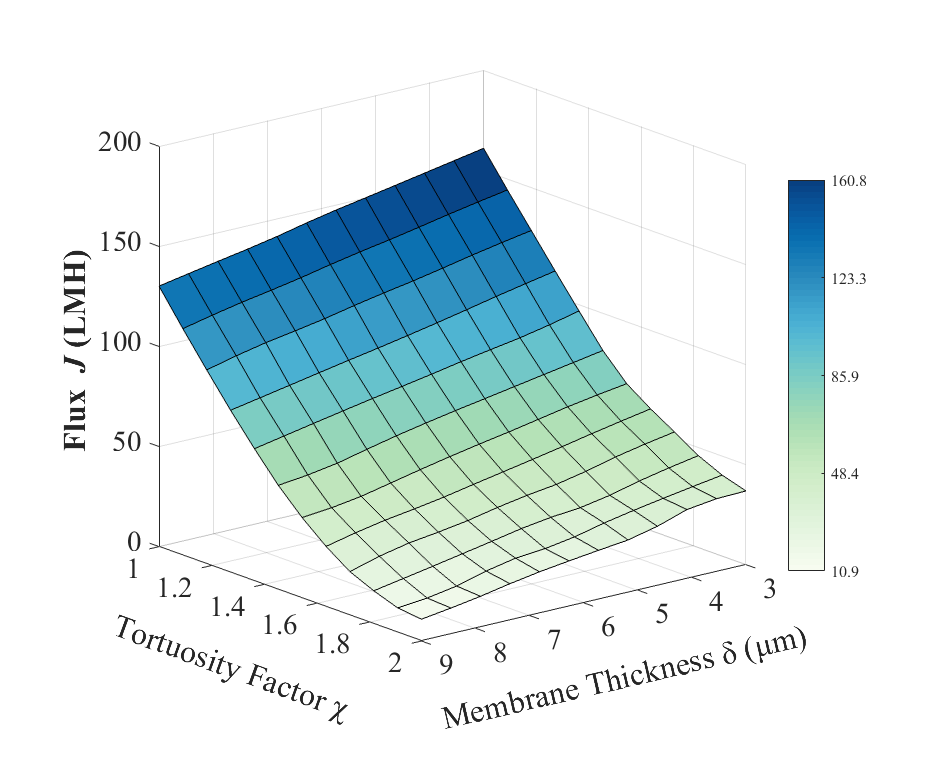}  
\renewcommand{\figurename}{\textnormal{\textbf{Figure}}}
    \caption{\replaced{\rmfamily Response surface plot of permeate flux as a function of membrane thickness and tortuosity factor.}{Response surface plot of temperature polarization as a function of membrane thickness and tortuosity factor.}}  
    \label{fig:example}  
\end{figure}

Figure 15 shows a 3D surface plot illustrating the effect of membrane thickness and pore size on permeate flux, and Figure 16 shows the projection. The results show that as pore size increases and membrane thickness decreases, the mass transfer resistance decreases, which increases permeate flux. The figure also shows that variations in membrane thickness have a more pronounced effect on permeate flux than changes in pore size, highlighting the importance of optimizing thickness as a primary design consideration.
\begin{figure}[htbp]
    \centering
\includegraphics[width=0.45\textwidth]{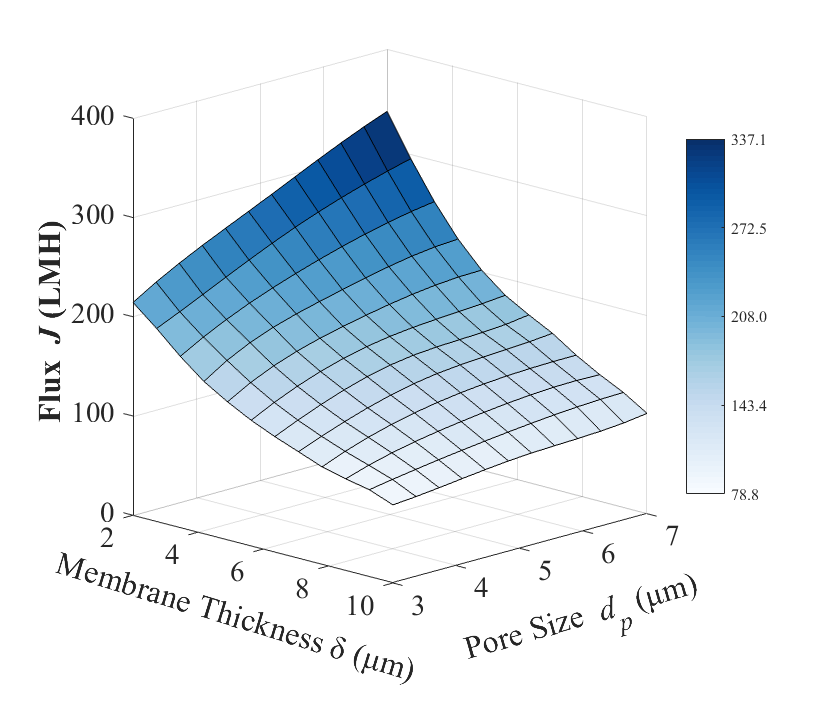}  
\renewcommand{\figurename}{\textnormal{\textbf{Figure}}}
    \caption{\replaced{\rmfamily Response surface plots of permeate flux as a function of membrane thickness and pore size.}{Response surface plots of temperature polarization as a function of membrane thickness and pore size.}}  
    \label{fig:example}  
\end{figure}

\begin{figure}[h]
    \centering
\includegraphics[width=0.45\textwidth]{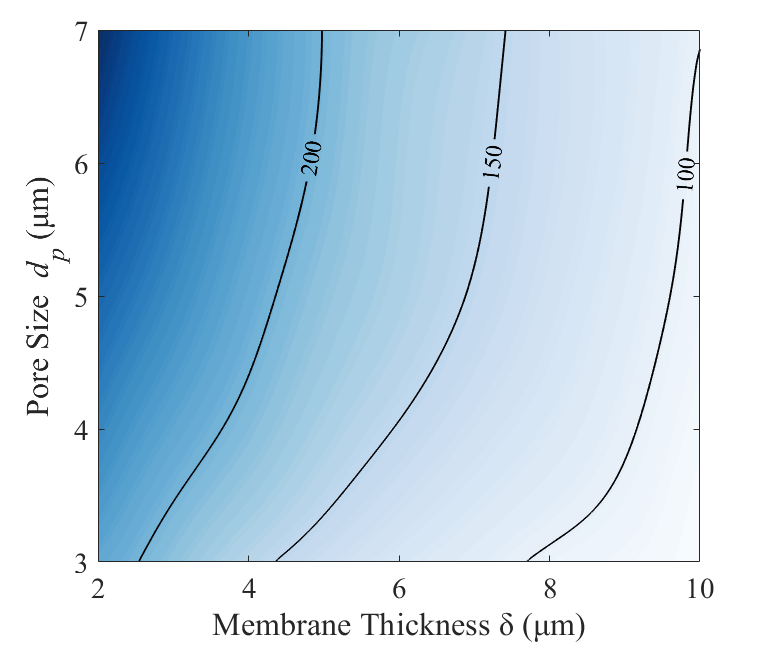}  
\renewcommand{\figurename}{\textnormal{\textbf{Figure}}}
    \caption{\replaced{\rmfamily The visualization demonstrates that as membrane thickness decreases and pore size increases, the permeate flux significantly rises.}{The visualization demonstrates that as membrane thickness decreases and pore size increases, the permeate flux significantly rises.}}  
    \label{fig:example}  
\end{figure}

Figure 16 illustrates the effect of membrane thickness and pore size on permeate flux, as shown by the projection. This projection effectively visualizes the relationship between these two critical parameters and their combined influence on permeate flux, enhancing our understanding of how variations in both thickness and pore size interact to affect the overall performance of the membrane.
This study provides a comprehensive investigation of the effects of several key parameters and the operating condition on the permeate flux in the VMD process, and the microstructural characteristics of the membrane. By comparing the experimental data of the novel PTEP membrane, the accuracy of the current model was validated. An in-depth analysis of the interactions between these parameters and their collective effects on membrane performance was performed.

\section{Conclusion}
First and foremost, the results show that higher membrane porosity and optimal membrane thickness significantly increase the permeate flux. This is attributed to the increased evaporation surface area provided by higher porosity, which facilitates the transfer of water vapor. \added{The optimized membrane thickness effectively reduces the mass transfer resistance, and the synergistic effect of these two factors significantly increases the permeate flux efficiency.} 
Further investigation reveals that membrane wettability of the membrane and the pore shape play a critical role in the diffusion capacity of gases. 
In particular, as temperatures rise, membranes with good wettability can effectively prevent pore wetting and ensure smooth vapor flow. \added{This finding highlights the critical importance of membrane surface properties in high temperature operating environments.}
At last, the ranking of the effects of different parameters on permeate flux was established through experimental investigations. \added{The results showed that feed temperature and tortuosity factor have comparable magnitudes of influence on permeate flux, as indicated by their similar slopes in the three-dimensional analysis. Both factors are more influential than membrane thickness, and membrane thickness is more influential than pore size.}\deleted{The results showed that feed temperature is approximately equal to the tortuosity factor, which is more influential than membrane thickness, and membrane thickness is more
influential than pore size.} 
This ranking not only highlights the dominant role of feed temperature and tortuosity factor in influencing membrane performance, but also provides guidance for subsequent membrane design and optimization. 
\deleted{In addition, the visualizing of the complex relationships between various parameters provides theoretical support for optimizing membrane material design and increasing the efficiency of VMD systems. 
The response surface plots effectively illustrate the influence of tortuosity factor, membrane thickness, and temperature on permeate flux, highlighting the critical role of feed temperature in membrane design. This finding opens new avenues for membrane material development.}
\added{In summary, this study combines membrane characteristics with operating conditions through a hierarchical optimization strategy, providing a new theoretical framework for the design of high-performance membrane materials. This finding not only deepens the understanding of the coupling effects of multiple parameters, but also lays a solid foundation for the practical application of vacuum membrane distillation systems.}

Future research will focus on translating these theoretical findings into practical applications to address issues related to water scarcity and environmental pollution. We recommend further exploration of novel membrane materials and improved membrane fabrication techniques to achieve more efficient VMD systems. Additionally, experimental validation under specific operating conditions will be an important direction for future studies to ensure the reliability and applicability of theoretical models. In conclusion, this study provides a solid foundation for the further development of membrane evaporation technology by demonstrating the potential to improve performance by fine-tuning the physicochemical properties of membranes. This not only provides new perspectives for industrial applications, but also offers viable technological pathways to address the global water crisis.
	
\section{Acknowledgements}
This work was supported by the National Natural Science Foundation of China (Grant Nos. 12272100, 12462025), Guangxi Natural Science Foundation (Grant No. 2024JJA1\\10126) and Innovation Project of Guangxi Graduate Education (Grant No. XYSCR2024103), Guangxi “Bagui Scholar” Teams for Innovation and Research Project, and Guangxi Collaborative Innovation Center of Multisource Information Integration and Intelligent Processing.
	

	\bibliography{refs}

\begin{thebibliography}{10}
\expandafter\ifx\csname url\endcsname\relax
  \def\url#1{\texttt{#1}}\fi
\expandafter\ifx\csname urlprefix\endcsname\relax\def\urlprefix{URL }\fi
\expandafter\ifx\csname href\endcsname\relax
  \def\href#1#2{#2} \def\path#1{#1}\fi

\bibitem{Chamani2021}
H.~Chamani, J.~Woloszyn, T.~Matsuura, D.~Rana, C.~Q. Lan, Pore wetting in membrane distillation: A comprehensive review, Progress in Materials Science 122 (2021) 100843.

\bibitem{suleman2021temperature}
M.~Suleman, M.~Asif, S.~A. Jamal, Temperature and concentration polarization in membrane distillation: A technical review, Desalination and Water Treatment 229 (2021) 52--68.

\bibitem{qasim2019reverse}
M.~Qasim, M.~Badrelzaman, N.~N. Darwish, N.~A. Darwish, N.~Hilal, Reverse osmosis desalination: A state-of-the-art review, Desalination 459 (2019) 59--104.

\bibitem{el2006framework}
M.~S. El-Bourawi, Z.~Ding, R.~Ma, M.~Khayet, A framework for better understanding membrane distillation separation process, Journal of membrane science 285~(1-2) (2006) 4--29.

\bibitem{qtaishat2008heat}
M.~Qtaishat, T.~Matsuura, B.~Kruczek, M.~Khayet, Heat and mass transfer analysis in direct contact membrane distillation, Desalination 219~(1-3) (2008) 272--292.

\bibitem{khayet2011membranes}
M.~Khayet, Membranes and theoretical modeling of membrane distillation: A review, Advances in colloid and interface science 164~(1-2) (2011) 56--88.

\bibitem{sholl2016seven}
D.~S. Sholl, R.~P. Lively, Seven chemical separations to change the world, Nature 532~(7600) (2016) 435--437.

\bibitem{zare2018membrane}
S.~Zare, A.~Kargari, Membrane properties in membrane distillation, in: Emerging technologies for sustainable desalination handbook, Elsevier, 2018, pp. 107--156.

\bibitem{niknejad2021high}
A.~S. Niknejad, S.~Bazgir, A.~Kargari, M.~Barani, E.~Ranjbari, M.~Rasouli, A high-flux polystyrene-reinforced styrene-acrylonitrile/polyacrylonitrile nanofibrous membrane for desalination using direct contact membrane distillation, Journal of Membrane Science 638 (2021) 119744.

\bibitem{izquierdo2003factors}
M.~A. Izquierdo-Gil, G.~Jonsson, Factors affecting flux and ethanol separation performance in vacuum membrane distillation (vmd), Journal of Membrane Science 214~(1) (2003) 113--130.

\bibitem{banat1999modeling}
F.~A. Banat, F.~A. Al-Rub, M.~Shannag, Modeling of dilute ethanol--water mixture separation by membrane distillation, Separation and Purification Technology 16~(2) (1999) 119--131.

\bibitem{a2015review}
M.~M. A~Shirazi, A.~Kargari, A review on applications of membrane distillation (md) process for wastewater treatment, Journal of Membrane Science and Research 1~(3) (2015) 101--112.

\bibitem{zakrzewska1999concentration}
G.~Zakrzewska-Trznadel, M.~Harasimowicz, A.~G. Chmielewski, Concentration of radioactive components in liquid low-level radioactive waste by membrane distillation, Journal of Membrane Science 163~(2) (1999) 257--264.

\bibitem{li2023enhanced}
J.~Li, C.~Xu, J.~Ye, E.~Li, S.~Xu, M.~Huang, Enhanced anti-fouling of forward osmosis membrane by pulsatile flow operation in textile wastewater treatment, Desalination 565 (2023) 116878.

\bibitem{niknejad2021mechanically}
A.~S. Niknejad, S.~Bazgir, A.~Kargari, Mechanically improved superhydrophobic nanofibrous polystyrene/high-impact polystyrene membranes for promising membrane distillation application, Journal of Applied Polymer Science 138~(36) (2021) 50917.

\bibitem{weyl1967recovery}
P.~K. Weyl, Recovery of demineralized water from saline waters, uS Patent 3,340,186 (Sep.~5 1967).

\bibitem{olatunji2018heat}
S.~O. Olatunji, L.~M. Camacho, Heat and mass transport in modeling membrane distillation configurations: a review, Frontiers in Energy Research 6 (2018) 130.

\bibitem{shokrollahi2020producing}
M.~Shokrollahi, M.~Rezakazemi, M.~Younas, Producing water from saline streams using membrane distillation: modeling and optimization using cfd and design expert, International Journal of Energy Research 44~(11) (2020) 8841--8853.

\bibitem{qi2020numerical}
J.~Qi, J.~Lv, Z.~Li, W.~Bian, J.~Li, S.~Liu, A numerical simulation of membrane distillation treatment of mine drainage by computational fluid dynamics, Water 12~(12) (2020) 3403.

\bibitem{ding2023enhancement}
X.~Ding, F.~Wang, G.~Lin, B.~Tang, X.~Li, G.~Zhou, W.~Wang, J.~Zhang, Y.~Shi, The enhancement of separation performance of hollow fiber membrane modules: From the perspective of membranes and membrane modules structural optimization design, Chemical Engineering Science 280 (2023) 119106.

\bibitem{teoh2009membrane}
M.~M. Teoh, T.-S. Chung, Membrane distillation with hydrophobic macrovoid-free pvdf--ptfe hollow fiber membranes, Separation and Purification Technology 66~(2) (2009) 229--236.

\bibitem{shan1993lattice}
X.~Shan, H.~Chen, Lattice boltzmann model for simulating flows with multiple phases and components, Physical review E 47~(3) (1993) 1815.

\bibitem{li2013lattice}
Q.~Li, K.~Luo, X.~Li, Lattice boltzmann modeling of multiphase flows at large density ratio with an improved pseudopotential model, Physical Review E—Statistical, Nonlinear, and Soft Matter Physics 87~(5) (2013) 053301.

\bibitem{kruger2017lattice}
T.~Kr{\"u}ger, H.~Kusumaatmaja, A.~Kuzmin, O.~Shardt, G.~Silva, E.~M. Viggen, The lattice boltzmann method, Springer International Publishing 10~(978-3) (2017) 4--15.

\bibitem{li2015lattice}
Q.~Li, Q.~Kang, M.~M. Francois, Y.~He, K.~Luo, Lattice boltzmann modeling of boiling heat transfer: The boiling curve and the effects of wettability, International Journal of Heat and Mass Transfer 85 (2015) 787--796.

\bibitem{shan1994simulation}
X.~Shan, H.~Chen, Simulation of nonideal gases and liquid-gas phase transitions by the lattice boltzmann equation, Physical Review E 49~(4) (1994) 2941.

\bibitem{qin2024efficient}
Z.~Qin, X.~Lu, L.~Lv, Z.~Tang, B.~Wen, An efficient gpu algorithm for lattice boltzmann method on sparse complex geometries, IEEE Transactions on Parallel and Distributed Systems (2024).

\bibitem{suk2010development}
D.~E. Suk, T.~Matsuura, H.~B. Park, Y.~M. Lee, Development of novel surface modified phase inversion membranes having hydrophobic surface-modifying macromolecule (nsmm) for vacuum membrane distillation, Desalination 261~(3) (2010) 300--312.

\bibitem{urtiaga2000kinetic}
A.~Urtiaga, G.~Ruiz, I.~Ortiz, Kinetic analysis of the vacuum membrane distillation of chloroform from aqueous solutions, Journal of membrane science 165~(1) (2000) 99--110.

\bibitem{al2006concentration}
S.~Al-Asheh, F.~Banat, M.~Qtaishat, M.~Al-Khateeb, Concentration of sucrose solutions via vacuum membrane distillation, Desalination 195~(1-3) (2006) 60--68.

\bibitem{diban2009vacuum}
N.~Diban, O.~C. Voinea, A.~Urtiaga, I.~Ortiz, Vacuum membrane distillation of the main pear aroma compound: Experimental study and mass transfer modeling, Journal of Membrane Science 326~(1) (2009) 64--75.

\bibitem{criscuoli2008treatment}
A.~Criscuoli, J.~Zhong, A.~Figoli, M.~Carnevale, R.~Huang, E.~Drioli, Treatment of dye solutions by vacuum membrane distillation, Water Research 42~(20) (2008) 5031--5037.

\bibitem{bandini1997vacuum}
S.~Bandini, A.~Saavedra, G.~C. Sarti, Vacuum membrane distillation: experiments and modeling, AIChE journal 43~(2) (1997) 398--408.

\bibitem{sarti1993extraction}
G.~Sarti, C.~Gostoli, S.~Bandini, Extraction of organic components from aqueous streams by vacuum membrane distillation, Journal of membrane science 80~(1) (1993) 21--33.

\bibitem{srisurichan2006mass}
S.~Srisurichan, R.~Jiraratananon, A.~Fane, Mass transfer mechanisms and transport resistances in direct contact membrane distillation process, Journal of membrane science 277~(1-2) (2006) 186--194.

\bibitem{gryta1997membrane}
M.~Gryta, M.~Tomaszewska, A.~Morawski, Membrane distillation with laminar flow, Separation and Purification Technology 11~(2) (1997) 93--101.

\bibitem{gostoli1989separation}
C.~Gostoli, G.~Sarti, Separation of liquid mixtures by membrane distillation, Journal of Membrane Science 41 (1989) 211--224.

\bibitem{franken1987wetting}
A.~Franken, J.~Nolten, M.~Mulder, D.~Bargeman, C.~Smolders, Wetting criteria for the applicability of membrane distillation, Journal of membrane science 33~(3) (1987) 315--328.

\bibitem{qiu2019graphynes}
H.~Qiu, M.~Xue, C.~Shen, Z.~Zhang, W.~Guo, Graphynes for water desalination and gas separation, Advanced Materials 31~(42) (2019) 1803772.

\bibitem{yin2022ultrafast}
Y.~Yin, Y.~Yang, G.~Liu, H.~Chen, D.~Gong, Y.~Ying, J.~Fan, S.~Liu, Z.~Li, C.~Wang, et~al., Ultrafast solid-phase synthesis of 2d pyrene-alkadiyne frameworks towards efficient capture of radioactive iodine, Chemical Engineering Journal 441 (2022) 135996.

\bibitem{gong2024alkadiyne}
D.~Gong, B.~Wen, L.~Wang, H.~Zhang, H.~Chen, J.~Fan, Z.~Li, L.~Guo, G.~Shi, Z.~Zhu, et~al., Alkadiyne--pyrene conjugated frameworks with surface exclusion effect for ultrafast seawater desalination, Journal of the American Chemical Society 146~(5) (2024) 3075--3085.

\bibitem{aidun2010lattice}
C.~K. Aidun, J.~R. Clausen, Lattice-boltzmann method for complex flows, Annual review of fluid mechanics 42~(1) (2010) 439--472.

\bibitem{chen1998lattice}
S.~Chen, G.~D. Doolen, Lattice boltzmann method for fluid flows, Annual review of fluid mechanics 30~(1) (1998) 329--364.

\bibitem{li2016lattice}
Q.~Li, K.~H. Luo, Q.~Kang, Y.~He, Q.~Chen, Q.~Liu, Lattice boltzmann methods for multiphase flow and phase-change heat transfer, Progress in Energy and Combustion Science 52 (2016) 62--105.

\bibitem{chen2014critical}
L.~Chen, Q.~Kang, Y.~Mu, Y.-L. He, W.-Q. Tao, A critical review of the pseudopotential multiphase lattice boltzmann model: Methods and applications, International journal of heat and mass transfer 76 (2014) 210--236.

\bibitem{huang2015multiphase}
H.~Huang, M.~Sukop, X.~Lu, Multiphase lattice boltzmann methods: Theory and application (2015).

\bibitem{lallemand2000theory}
P.~Lallemand, L.-S. Luo, Theory of the lattice boltzmann method: Dispersion, dissipation, isotropy, galilean invariance, and stability, Physical review E 61~(6) (2000) 6546.

\bibitem{mccracken2005multiple}
M.~E. McCracken, J.~Abraham, Multiple-relaxation-time lattice-boltzmann model for multiphase flow, Physical Review E—Statistical, Nonlinear, and Soft Matter Physics 71~(3) (2005) 036701.

\bibitem{kupershtokh2009equations}
A.~L. Kupershtokh, D.~Medvedev, D.~Karpov, On equations of state in a lattice boltzmann method, Computers \& Mathematics with Applications 58~(5) (2009) 965--974.

\bibitem{shan1995multicomponent}
X.~Shan, G.~Doolen, Multicomponent lattice-boltzmann model with interparticle interaction, Journal of statistical physics 81 (1995) 379--393.

\bibitem{wen2017chemical}
B.~Wen, X.~Zhou, B.~He, C.~Zhang, H.~Fang, Chemical-potential-based lattice boltzmann method for nonideal fluids, Physical Review E 95~(6) (2017) 063305.

\bibitem{wen2020chemical}
B.~Wen, L.~Zhao, W.~Qiu, Y.~Ye, X.~Shan, Chemical-potential multiphase lattice boltzmann method with superlarge density ratios, Physical Review E 102~(1) (2020) 013303.

\bibitem{d1992generalized}
D.~d'Humieres, Generalized lattice-boltzmann equations, Rarefied gas dynamics (1992).

\bibitem{jamet2002theory}
D.~Jamet, D.~Torres, J.~Brackbill, On the theory and computation of surface tension: the elimination of parasitic currents through energy conservation in the second-gradient method, Journal of Computational Physics 182~(1) (2002) 262--276.

\bibitem{swift1995lattice}
M.~R. Swift, W.~Osborn, J.~M. Yeomans, Lattice boltzmann simulation of nonideal fluids, Physical review letters 75~(5) (1995) 830.

\bibitem{liu2023contact}
Y.~Liu, Y.~Yao, Q.~Li, X.~Zhong, B.~He, B.~Wen, Contact angle measurement on curved wetting surfaces in multiphase lattice boltzmann method, Langmuir 39~(8) (2023) 2974--2984.

\bibitem{lallemand2003theory}
P.~Lallemand, L.-S. Luo, Theory of the lattice boltzmann method: Acoustic and thermal properties in two and three dimensions, Physical review E 68~(3) (2003) 036706.

\bibitem{khayet2004characterization}
M.~Khayet, K.~Khulbe, T.~Matsuura, Characterization of membranes for membrane distillation by atomic force microscopy and estimation of their water vapor transfer coefficients in vacuum membrane distillation process, Journal of membrane science 238~(1-2) (2004) 199--211.

\end{thebibliography}

\end{document}